\documentclass[useAMS,usenatbib,letterpaper]{mn2e}
\usepackage{graphicx, widetext}
\usepackage{amssymb, amsmath,mydefs}
\usepackage{mathptm}
\usepackage{color}
\usepackage{lscape}
\usepackage[section]{placeins}
\usepackage{epstopdf}
\usepackage{epsf}
\usepackage{ulem}
\usepackage{hyperref}
\usepackage{array}
\usepackage{multirow}

\newcommand{\skdis}[1]{
\mathbf{s}^{r(#1)}(\mathbf{k})}

\newcommand{\sxdis}[1]{{\mathbf s}^{r(#1)}(\bx)}
\newcommand{\sxdiso}[1]{{\mathbf s}^{r(#1)}(\bx_0)}
\newcommand{\sxdist}[1]{{\mathbf s}^{r(#1)}(\bx_1)}
\newcommand{\sxdisn}[1]{{\mathbf s}^{r(#1)}(\bx_{n-1})}
\newcommand{\ssqdiso}[1]{{\mathbf s}^{(#1)}(\mathbf{q}_0)}

\newcommand{\sqdist}[1]{{\mathbf s}^{r(#1)}(\mathbf{q}_1)}
\newcommand{\sqdisn}[1]{{\mathbf s}^{r(#1)}(\mathbf{q}_{n-1})}

\newcommand{\ssdiso}[1]{{\mathbf s}^{(#1)}(\bx_0)}
\def\StdRec {\textbf{StdRec}}
\def\DisIter {\textbf{DisIter}}
\def\DisIterss {\textbf{DisIterSS}}
\def\StdIter {\textbf{StdIter}}
\def\StdIterss {\textbf{StdIterSS}}
\usepackage{mydefs}
\def\apj{\rm{ApJ}}
\def\apjl{\rm{ApJ}}
\def\apjs{\rm{ApJS}}
\def\aap{\rm{A\&A}}                
\def\jcap{\rm{JCAP}}                
\def\mnras{\rm{MNRAS}}             
\def\prd{\rm{Phys.~Rev.~D}}        
\def\nat{\rm{Nature}}              

\title[Iterative reconstruction for BAO and beyond]{Iterative reconstruction excursions for Baryon Acoustic Oscillations and beyond}
\author[Hee-Jong Seo \etal]{Hee-Jong Seo${}^1$\thanks{seoh@ohio.edu}. Atsuhisa Ota${}^1$, Marcel Schmittfull${}^2$, Shun Saito${}^{3,4}$,
Florian Beutler${}^5$
\\
${}^1$Department of Physics and Astronomy Ohio University, Athens, OH, 45701, USA\\
 ${}^2$ Institute for Advanced Study, Einstein Drive, Princeton, NJ 08540, USA\\
 ${}^{3}$ Institute for Multi-messenger Astrophysics and Cosmology, Department of Physics, Missouri University of Science and Technology, \\
                    1315 N. Pine St., Rolla MO 65409, USA\\
 ${}^{4}$Kavli Institute for the Physics and Mathematics of the Universe (WPI), Todai Institutes for Advanced Study,\\
                   the University of Tokyo, Kashiwanoha, Kashiwa, Chiba 277-8583, Japan\\
 ${}^{5}$Institute for Astronomy, University of Edinburgh, Royal Observatory, Blackford Hill, Edinburgh EH9 3HJ, UK}

\date{\today} 

\begin{document}

\maketitle

\begin{abstract}
The density field reconstruction technique has been widely used for recovering the Baryon Acoustic Oscillation (BAO) feature in galaxy surveys that has been degraded due to nonlinearities. 
Recent studies advocated adopting iterative steps to improve the recovery much beyond that of the standard technique. In this paper, we investigate the performance of a few selected iterative reconstruction techniques focusing on the BAO and the broadband-shape of the two-point clustering.
We include redshift-space distortions, halo bias, and shot noise and inspect the components of the reconstructed field in Fourier space and in configuration space using both density field-based reconstruction and displacement field-based reconstruction. We find that the displacement field reconstruction becomes quickly challenging in the presence of non-negligible shot noise and therefore present surrogate methods that can be practically applied to a much more sparse field such as galaxies. For a galaxy field, implementing a debiasing step to remove the Lagrangian bias appears crucial for the displacement field reconstruction. 
We show that the iterative reconstruction does not substantially improve the BAO feature beyond an aggressively optimized standard reconstruction with a small smoothing kernel. However, we find taking iterative steps allows us to use a small smoothing kernel more `stably', i.e., without causing a substantial deviation from the linear power spectrum  on large scales. 
In one specific example we studied, we find that a deviation of 13\% in $P(k\sim 0.1\ihMpc)$  with an aggressive standard reconstruction can reduce to 3-4\% with iterative steps.
\end{abstract}

\section{Introduction}

Baryon acoustic oscillations (hereafter BAO) from galaxy surveys have played a key role in today's cosmology inference on dark energy and the Hubble constant~\citep[e.g,][]{Aubourg:2015,ebossBAO}.  
The BAO feature was formed by primordial sound waves that propagated through tightly coupled photons and baryons in the very early Universe; they were subsequently frozen out at the epoch of recombination when the photons and baryons decoupled.
The observed sizes of BAO from galaxy surveys, in comparison to its true physical size estimated from an independent probe such as from the Cosmic Microwave Background (CMB), construct a robust standard ruler test that provides cosmological distances (i.e., angular diameter distances and Hubble parameters) as a function of time~\citep[e.g.,][]{SE:2003}. The signal of this primordial feature degrades with the structure growth of the Universe; nonlinearity associated with the structure growth decreases the precision and accuracy of the measurement~\citep[e.g,][]{Meiksin,Crocce:2008,Seo:2008,Matsubara:2008,Pad:2009,Seo:2010}. 

The density field reconstruction technique~\citep{Eisenstein:2017rec} has been widely used to recover the BAO feature in galaxy surveys from the effect of this structure growth. The performance of this method is quite stable and robust, straightforwardly depending on the interplay between the smoothing kernel and the signal to noise of the input data~\citep[e.g.,][]{White:2010}. A perturbation theory-based empirical model has provided a good description for the reconstructed BAO feature~\citep[e.g,][]{ESW07,Crocce:2008,Padmanabhan:2009Lag,Noh:2019wp,Seo:2016brs} and has been used for  the BAO-only analysis; the resulting constraints have been critical for understanding dark energy~\citep[e.g.,][]{ebossBAO}.

In addition to the BAO feature, large-scale structure contains other cosmology probes. The overall (i.e., broad-band) shape of the clustering provides information about the horizon scale at the epoch of the matter-radiation equality~\citep[e.g,][]{EH98}. The matter-radiation equality scale imprinted in the broad-band shape can, in principle, serve as another standard ruler to measure cosmological distance scales. Second, the underlying clustering of matter/galaxies should be isotropic.  A deviation from the isotropy provides information on the geometry of the Universe, which is called the Alcock-Paczynski (AP) test~\citep{AP}, and on the peculiar velocity field, which is called redshift-space distortions (RSD)~\citep{Kaiser:1987}. Features such as RSD and the broad-band shape are more prone to nonlinear effects because they are less distinct/localized than the BAO feature.

The standard density field reconstruction alters this shape of the broadband and RSD. Although the linear BAO information is largely recovered, the resulting power spectrum does not agree with the linear power spectrum, because the nonlinearity is not fully reversed even at the second order in density perturbations ~\citep[e.g,][]{Padmanabhan:2009Lag,Schmittfull:2015mja}.  Understanding the full shape after reconstruction will enable a full/combined clustering fit to the BAO and AP+RSD features using the post-reconstructed field. Due to the reduced nonlinearity, the perturbation theory (PT) may be in a better agreement with the post-reconstructed density field at a smaller scale and a lower redshift, compared to the raw density field.
A promising progress has been made in this direction in the recent literature. \citet{Hikage:2017real,Hikage:2020red} used the standard perturbation (PT) to 1-loop and derived a model for the post-reconstruction real-space and redshift-space matter power spectrum, showing that the PT model can indeed explain the reduced mode-coupling effect after reconstruction and is in a better agreement with the post-reconstruction clustering. \citet{White2015Zel} and \citet{Chen:2019rec} used the Zeldovich approximation to build both Fourier space and configuration space models for the post-reconstruction field, accounting for the redshift-space distortions as well as galaxy bias, demonstrating a good agreement between the model and the simulation. These studies show the PT models perform well in describing  the galaxy clustering even upto $k\sim 0.2-0.4\ihMpc$ at a moderate redshift. But as one attempts to extract a smaller scale information, which can potentially boost the reconstructed BAO information, it becomes more difficult for these theories to predict the nonlinear component of the information that propagates through the reconstruction operation and the agreement becomes worse.

Recently, a variety of iterative extensions to the BAO reconstruction technique have been developed, aimed at  maximally extracting linear BAO information from the observed nonlinear fields \citep[e.g.,][]{Tassev:2012,Schmittfull:2017,Hada2018,Hada2019,Zhu2017,Yu2017,Wang2017,Zhu2018}. The details of implementation are different from study to study, but all these methods attempt to iteratively reconstruct more accurate displacement field that mass tracers should have experienced and take the divergence of the estimated displacement field as the reconstructed density field.
As \citet{Baldauf:2015dis} showed, the dark matter displacement field at lower redshift is very highly correlated with the initial density field (e.g., correlation higher than $0.95$ for $k < 0.5\ihMpc$  at $z=0.6$ from \citet{Ota:2021}) as the shift term that is responsible for the BAO degradation is much smaller than in the nonlinear density field, implying almost a complete BAO information. In fact, this shift term contribution is not only small, but is positive in the case of the displacement field, while it is negative (i.e., BAO damping) in the nonlinear density field. These iterative reconstruction methods accordingly show substantially improved BAO feature compared to the standard reconstruction, at least in the presence of very little shot noise. 
Note that, if the nonlinear displacement is faithfully recovered from this reconstruction, its divergence field will not be the same as the linear field, but will still contain most of the BAO information \citep[e.g.,][]{Baldauf:2015dis, Ota:2021}.

We can consider this iterative operation as solving the nonlinear equation for the nonlinear displacement field with a linear operation on the filtered nonlinear density field, while iteratively correcting the smaller scale displacement until the final field becomes uniform/Lagrangian~\citep{Schmittfull:2017}. The standard reconstruction on the other hand is known to perform sub-optimally when this filtering kernel is reduced much smaller than $\sim 7\hMpc$~\citep[e.g.,][]{Seo:2016brs}.
Although the iterative operation is more complex than the simple, standard reconstruction, its final product could be therefore more stable on small scales, i.e., be able to recover a much smaller scale information without suffering the performance degradation. Comparing the broadband shape of the iteratively reconstructed field with the corresponding perturbation theory model, if such model could be constructed, would be quite useful for optimizing both the post-reconstruction BAO and broadband analysis. To address this point, our companion paper, \citet{Ota:2021} constructs a 1-loop perturbation theory model for the method in \citet{Schmittfull:2017} for mass tracers in real space and compare with the simulations.

In this paper, we numerically investigate the properties of iterative reconstructions on the BAO and also the broadband, focusing on the iterative implementation presented in \citet{Schmittfull:2017}. We focus on two iterative reconstruction schemes in that study that closely follow the method of the standard reconstruction scheme. The first method, noted as `O(1)' in their paper, iteratively operates to estimate the nonlinear displacement field. The second method, `the iterative standard reconstruction', was  originally tested in ~\citet{Seo:2010},  but modified by \citet{Schmittfull:2017}. While the latter paper focused on the real-space, almost shot-noiseless dark matter field, we extend the test to the redshift space and the tracers with halo bias and inspect the components of the reconstructed field in Fourier space and in configuration space. In the process of extension, we invent a practical surrogate method for  \citet{Schmittfull:2017} that can be easily applied to the galaxy field with high shot noise. 
We inspect and compare different reconstructions mainly using three indicators: propagators as an indicator of the shift term contribution  (or the BAO damping), cross-correlation coefficients as an indicator of the residual mode coupling contribution and a signal to noise, and the shape of the power spectrum and correlation function as the combination of all terms. We compare the iterative reconstruction with the standard reconstruction at the optimal case of each method.
We will define the performance in terms of how well the reconstructed field is correlated with the initial density field. Also given the lack of theory model to compare with, we will define the performance based on how close the overall clustering of the final field is close to that of the initial field.

The structure of the paper is as follows. In \S~\ref{sec:methods}, we explain the implementation of the iterative reconstructions for redshift space and for the biased tracers. In \S~\ref{sec:Results}, we present the results. 
Finally, in \S~\ref{sec:conclusion}, we summarize our results.

\section{Methods}\label{sec:methods}
The four methods of iterative reconstructions we test/develop in this paper (\StdIter, \StdIterss, \DisIter, \DisIterss\ as summarized in  Tab.~\ref{tab:method}) closely follow the standard reconstruction that we describe below.

\subsection{Reconstruction Methods}
\subsubsection{The Standard reconstruction: \StdRec}\label{subsubsec:firststep}

We describe the process of the standard reconstruction developed in \citet{Eisenstein:2017rec}, while adopting the convention `Rec-Sym' in \citet{White2015Zel} and \citet{Chen:2019rec}~\footnote{The same convention was called `Rec-Cohn' in \citet{Ding2018}}for treating the redshift-space distortions. This is also the first step of the iterative reconstruction. 

We first start with the observed nonlinear density field of matter or galaxies, $\tilde{\delta}^s_{\rm NL}$(\bx), in observed location in $\bx$, apply the continuity equation to estimate the displacement field. To ensure the continuity equation to be valid against the effect of shot noise and the nonlinearity on small scales, we use a smoothing Gaussian Kernel, $S(\bk)$. While $S(\bk)$ can be chosen to be anisotropic, we choose the isotropic form as our default:
\begin{align}
S(\bk)  =  \exp\left( -\frac{k^2\Sigma^2}{4}\right).
\end{align}
$\Sigma$ is the smoothing scale and we note the definition of the smoothing scale varies in the literature typically by $\sqrt{2}$. Later, we will decrease the smoothing scale gradually during iteration for iterative reconstruction 

The resulting real-space displacement estimator can be written in Fourier space as:

\begin{eqnarray}
&&\skdis{0}{} = - \frac{i\bk}{k^2} \frac{\tilde{\delta}^s_{\rm NL}(\bk)}{b(1+\beta\mu^2)}S^{(0)}(\bk)\;,\nonumber\\
\label{eq:recisoq}
\end{eqnarray}
where the superscript `$(0)$' stands for the quantities before reconstruction. Here $\beta$ is the redshift-space distortion parameter,  $f$/b, where $f$ is the growth rate and $b$ is the galaxy bias, and $\mu$ is the cosine angle between the line of sight and $\vec{k}$. When $\tilde{\delta}^s_{\rm NL}$ is a real-space observable, $\beta$ is set to zero. 

The displacement field in configuration space ${\bx}$ is then derived by Fourier-transforming $\skdis{0}$;
\begin{eqnarray}
&&\skdis{0} \xrightarrow{\text{Fourier Transform}} \sxdis{0}.
\end{eqnarray}
The galaxies and the reference particles are displaced and their positions are updated. 
\begin{eqnarray}
&&\mbox{Galaxies: } \bx_1 = \bx_0+\ssdiso{0}\\
&&\mbox{where }\ssdiso{0} =  \sxdiso{0} + f (\mathbf{s}^{r(0)}\cdot\hat{\bz})\hat{\bz}, \mbox{ and}\\
&&\mbox{Reference:} \bq_1 = \bq_0+\ssqdiso{0}, 
\end{eqnarray}
where $f$ is the growth rate to account for the additional line-of-sight displacement due to redshift-space distortions, $\bx_0$ means the observed nonlinear location for the galaxies and $\mathbf{q}_0$ means the initial uniform location of the reference particles.  

The two density fields $\delta_{d}^{(1)}$ of galaxies and $\delta_{s}^{(1)}$ of reference particles are derived, respectively, and 
the reconstructed density field after the first reconstruction is then
\begin{eqnarray}
\delta_{\rm rec}^{(1)}(\bx_1) = \delta_{d}^{(1)} -\delta_{s}^{(1)}.
\end{eqnarray}
In this paper, we use \StdRec\ to denote the standard reconstruction.

\begin{table*}
\centering
\caption{\label{tab:method} Summary of the reconstruction methods investigated in this paper. The last column $\delta_{\rm rec}$ shows how the reconstructed clustering is defined in each case.}
\begin{tabular}{c|ccccccc}
\hline\hline
 Name & Description  & Iteration & Tracers & $\delta_{\rm rec}$ &Pixel window function\\ & & &&& and shot noise  \\\hline
 \StdRec & Standard reconstruction (\S~\ref{subsubsec:firststep}) & No & galaxies and reference field & $\delta_d -\delta_s$ & Corrected \\
 \StdIter & Standard Iterative reconstruction (\S~\ref{subsec:StdIter}) & Yes & galaxies and reference field & $\delta_d -\delta_s$ & Corrected \\
 \StdIterss & Single-field Standard Iterative reconstruction (\S~\ref{subsec:StdIterSS}) & Yes & reference field  & $\delta_s$ & Corrected\\
 \DisIter & Iterative displacement reconstruction (\S~\ref{subsec:DisIter}) & Yes & galaxies & $\nabla \cdot \mathbf{\chi}$ & Not corrected \\
 \DisIterss &  Reference-field iterative displacement reconstruction (\S~\ref{subsec:DisIterSS}) & Yes & reference field & $\nabla \cdot \mathbf{\chi}$ & Not corrected \\
\hline\hline
\end{tabular}
\end{table*}

\subsubsection{Standard iterative reconstruction: \StdIter}\label{subsec:StdIter}
The iterative methods we test here can be classified into two types. The first type is an extension of the standard reconstruction by adding iterative steps, as described in this section. 
This iterative reconstruction corresponds to `extended standard reconstruction' scheme defined in \citet{Schmittfull:2017}. It is based on the iterative reconstruction tested in \citet{Seo:2010} with a few modifications, mainly decreasing the smoothing scale in consecutive iterations steps. \citet{Schmittfull:2017} found this modification made a major difference in the performance.

We will expand the steps described in \S~\ref{subsubsec:firststep} by iteratively reducing the smoothing scale. Since any remaining small scale information we want to extract must be present in ${\delta}_{d}^{(1)}(\bx_1)$, we apply the continuity equation on $\delta_{d}^{(1)}(\bx_1)$ and derive 
\begin{eqnarray}
&&\skdis{1} = - \frac{i\bk}{k^2} {\delta}_{d}^{(1)}(\bk)S^{(1)}(\bk)\nonumber
\end{eqnarray}
 for matter. 
There are a few differences compared to the first step (i.e., the standard) reconstruction.
\begin{itemize}
\item Since we already have taken into account the most of the anisotropy in calculating $\delta_{d}^{(1)}$, we do not include the corresponding corrections after the first reconstruction. Including the anisotropy correction for higher iterations slightly reduces the performance along the line of sight for some cases, compared to what we present in this paper.  The bias correction after the first reconstruction is discussed in \S~\ref{subsec:methodbias}.
 \item Since the residual field ${\delta}_{d}^{(1)}(\bk)$ is mainly confined in small scales with reduced nonlinearity, $\skdis{1}$ would be incremental to  $\skdis{0}$.  
\item $S^{(1)}$ is using a smaller damping scale $\Sigma_1$ than $\Sigma_0$ for $S^{(0)}$. By default, we decrease the smoothing scale continuously by $\sqrt{2}$ in this paper to inspect the limit of reconstruction. In contrast, \citet{Schmittfull:2017} sets a minimum $\Sigsm$ so that the smoothing scale does not decrease when it reaches this minimum scale even though iteration continues. If we set the minimum smoothing scale, we find that the reconstruction result converges once the smoothing scale reaches that minimum scale as shown in ~\citet{Ota:2021}.  
\end{itemize}

After the second step reconstruction, the particles will be displaced as follows:
\begin{align}
\mbox{galaxies}: {\bx_{2}} = {\bx_{1}}+\sxdist{1}= \bx_0+\ssdiso{0}+\sxdist{1}\label{eq:xdisgal}\\
\mbox{references}: {\bf q_{2}} = {\bf q_{1}}+\sqdist{1}=\bq_0+\ssqdiso{0}+\sqdist{1}.
\end{align}
The density field of galaxies $\delta_d^{(2)}$ will be updated based on Eq.~\ref{eq:xdisgal} and ${\mathbf s^r(\bk)}$ for the next iteration will be derived. After the $n$-th reconstruction,
\begin{eqnarray}
&&\skdis{n} = - \frac{i\bk}{k^2} {\delta}_{d}^{(n)}(\bk)S^{(n)}(\bk).\nonumber
\end{eqnarray}
Note that $\sxdis{n}$ is always estimated from the density field of the (displaced) galaxies ${\delta}_{d}^{(n)}$ without using $\delta_{s}^{(n)}$. 
After  $n$-th reconstruction, the displaced particles will end up in the following position:

\begin{align}
\bx_{n} = \bx_0+\ssdiso{0}+\sxdist{1}+...+\sxdisn{n-1}\\
=   \bx_0 + \mathbf{s}_{net}(\bx_0),\label{eq:xdis}
\end{align}
where 
\begin{align}
\mathbf{s}_{net}(\bx_0) = \ssdiso{0}+\sxdist{1}+...+\sxdisn{n-1},\label{eq:xipo}
\end{align}
and the displaced reference particles will be located in
\begin{align}
\bq_{n} = \bq_0+\ssqdiso{0}+\sqdist{1}+...+\sqdisn{n}\\
=   \bq_0 + \mathbf{s}_{net}(\bq_0).\label{eq:xdisq}
\end{align}
We then evaluate 
\begin{eqnarray}
\delta_{\rm rec}^{(n)} = \delta_{d}^{(n)} -\delta_{s}^{(n)}.
\end{eqnarray}

For simplicity, we will refer to this method as `\StdIter' and we note that this method involves two fields, i.e., the displaced galaxy fields and the displaced reference fields. The final outcome is the difference of the two density fields just like the standard reconstruction, and therefore we can consider this as a density field reconstruction.

We examine the power spectra of $\delta_{d}^{(n)}$ and $\delta_{s}^{(n)}$ separately to understand the information transfer between the two fields during iteration, and they are referred to as the DD field and the SS field, respectively. 

In the following section, we will compare the performance of such iterative reconstruction for dark matter and biased tracers as a function of $n$ in comparison to the standard reconstruction (i.e. $n=1$). Again, we will define the performance in terms of how well the reconstructed field is correlated with the initial density field. Also given the lack theory model to compare with, particularly in the redshift space and with galaxy bias, we will define the performance based on how close the overall clustering of the final field is close to that of the initial field.

\subsubsection{A Single-field Standard Iterative Reconstruction: \StdIterss}\label{subsec:StdIterSS}
The standard iterative reconstruction explained in \ref{subsec:StdIter} gradually transfers the information from the displaced galaxy fields to the displaced reference fields, as will be shown in \S~\ref{sec:Results}.  Since we do not introduce the minimum smoothing scale, after a large number of iterations, \StdIter\ starts to degrade, but we find that the density field of the displaced {\it reference field alone}, $\delta_{s}^{n}$, continuously improves in terms of its correlation with the initial field, showing a more stable convergence behaviour than \StdIter. 
That is, at this limit, the power spectrum of $\delta_{s,I}$ alone is sufficiently reconstructed and one could choose to use only the reconstructed reference field for our cosmology analysis. We call this surrogate method `\StdIterss'.
\begin{table*}
\centering
\caption{\label{tab:sim} \Nbody\ Simulations used for different analyses. We utilize multiple sets of simulations, instead of a single consistent set, to inspect different aspects of reconstructed clustering due to our limited computational resources. All used a flat $\LCDM$ cosmology based on \citet{2016A&A...594A..13P} with $\Om = 0.3075$, $\Obhh=0.0223$, $h=0.6774$, and $\sigma_8=0.8159$. The last column, `Purpose' describes for which aspect each simulation was used. `Original mesh' is the grid size used for the force calculation and `FFT mesh' is the grid used for reconstruction and calculating the clustering statistics. For a halo field, we use two halo catalogs from the FastPM mock~\citep{Ding2018}, mainly due to a simulation availability, and we note the number density and the corresponding bias `b' for the two catalogs. }
\begin{tabular}{c|cccccccc}
\hline\hline
 & $z$  &  $L_{\rm box}(\hMpc)$ & $N_{\rm sim}$ & Original mesh & FFT mesh & $1/n(\itrihMpc)$ & Purpose \\\hline
 L500 & 0.6 & 500 & 5 & $1536^3$ & $512^3$ & 0.86  & Matter, lowest shot noise case.\\
 & & & & & & & Fig.~\ref{fig:DM500rkreal},\ref{fig:DM500realssdd},\ref{fig:DM500rkred},\ref{fig:DM500redcomp},\ref{fig:DMrsdSingle}\\
 subL500 &  &   &   & &  & 23.00  & Matter, increased shot noise\\
  & & & & & & & Fig.~\ref{fig:DM500redcompsub}\\
L1500 & 0.6 & 1500 & 1 & $1536^3$ & $1024^3$ &  23.28 & Matter with BAO and without BAO,\\
& & & & & & & BAO feature, Fig.~\ref{fig:DMPwPnw},\ref{fig:xiiter}\\
FastPM & 1.0 & 1350 & 1 &$4096^3$ &  $512^3$ & 817.45 (b=1.88), 265.44 (b=1.48) & Halos, high shot noise\\
& & & & & & & Fig.~\ref{fig:biased},\ref{fig:biasedssdd},\ref{fig:biasstdrec},\ref{fig:biasedm35}\\
\hline\hline
\end{tabular}
\end{table*}
\subsubsection{Iterative Displacement Reconstruction: \DisIter}\label{subsec:DisIter}
The second type of iterative reconstruction in this paper is reconstructing the displacement field itself, following \citet{Schmittfull:2017}, while there are a few minor differences in the setup.  This method therefore can be classified as a displacement field reconstruction. 
In particular, we extend the iterative method `{\it O(1)}' of \citet{Schmittfull:2017} that has been tested with real-space dark matter particles, to redshift space and biased tracers. Unlike \StdIter\ that evaluates the density fields of the displaced particles, this method evaluates the displacement field at the final position, i.e., at the estimated Lagrangian location of the observed galaxies. 

The procedure of iteration is the same upto Eq. \ref{eq:xipo}. The difference is that, after $n$-th reconstruction, we evaluate the divergence of $\mathbf{s}_{\rm net}(\bx_{n})$. It is crucial to evaluate $\mathbf{s}_{\rm net}$ at its final position $\bx_{n}$ not at its original, observed position $\bx_0$ (i.e., Eulerian position). The latter approximately returns the observed nonlinear density field.  The same aspect also makes this process different from estimating a later time velocity divergence field, where one evaluates at the Eulerian position.

In order to estimate the reconstructed displacement field $\mathbf{s}_{\rm net}(\bx_{n})$, we use the mass-weighted scheme. I.e., we collect all particles that ended up in a given mesh after applying the Cloud-in-Cell (CIC) assignment, and derive the mass weighted sum of the displacement for each pixel/mesh centered at $\bx_p$, $\hat{\mathbf{\chi}}_{\rm net}(\bx_p)$:

\def \bx{\mathbf x}
\begin{align}
	\hat{\mathbf{\chi}}_{\rm net}(\bx_p)
	&= \frac{\sum_i W_{\rm CIC}({\bx_p,\bx_i}) \mathbf{s}_{\rm net}(\bx_i)  }{\sum_i W_{\rm CIC}({\bx_p,\bx_i}) }
	,\label{eq:massdis}
\end{align}
where 
$ \mathbf{s}_{\rm net}(\bx_i)$ is the reconstructed displacement vector of the $i$-th particle, and $W_{\rm CIC}$ is the pixel window function indicating that we are using the Cloud-in-Cell assignment. For pixels with no particles found, we incorrectly set $\hat{\mathbf{\chi}}_{\rm net}(\bx_p)=0$.

The reconstructed field is then evaluated as the divergence of $\hat{\mathbf{\chi}}_{\rm net}$. In the Fourier space, 
\begin{align}
\delta_{\rm rec}^{(n)}(\bk ) = i\bk\cdot \hat{\mathbf{\chi}}_{\rm net}(\bk). 
\end{align}
As a difference from \citet{Schmittfull:2017}, we do not truncate the modes for $k$ greater than some maximum $k$.

The procedure is the same for the real space and the redshift space, except that $f=\beta=0$ in \S~\ref{subsubsec:firststep} in real space.

\subsubsection{Dealing with sparsity and the pixel window effect}\label{subsubsec:scarcity}
This method can be extended to biased tracers as biased tracers would have experienced the same displacement field as the matter except for on very small scales where the internal motion of halos begins to matter.  In reality, the biased tracers are often in a much smaller number and as a result many pixels of the field are empty without displacement tracers. To mitigate this effect, \citet{Schmittfull:2017} pads empty pixels with randomly chosen nearly, non-empty pixels. However, this procedure becomes increasingly inefficient with a decreasing number density: for biased tracers, most of the pixels would be empty if we set up the size of each pixel to be e.g., $5\hMpc$ for a very dense population with the number density of $0.001\trihMpc$. In this paper, we therefore do not pad the empty pixels. 

The problem of missing displacement tracers appears more tricky to deal with than missing tracers in the density field. First, the overall amplitude is reduced by the fraction of the zero-ed pixels, just like the effect of the survey window function. We correct for this with a simple multiplicative rescaling of the amplitude by the ratio of the periodic box volume to the effective volume traced by non-empty pixels~\citep[e.g., eq. 21 of][]{Peacock:1999}. Also, the sparsity introduces a large spurious power at large $k$, as will be shown in \S~\ref{subsubsec:shotnoise}. We try to mitigate such effect using a surrogate model \DisIterss\ or/and applying {\it debiasing} (\S~\ref{subsec:methodbias}), as explained below, when the sparsity becomes an issue.   

At the limit of one particle per mesh, i.e., with almost no empty pixels, we find that the mass-weighted displacement field from Eq.~\ref{eq:massdis} has the ordinary CIC pixel window function effect largely cancelled out between the numerator and the denominator, but there is a residual effect at the level of 1\% at $k\sim 0.2\ihMpc$ with our FFT mesh resolution \citep{Ota:2021}. With more empty pixels, this pixel window function effect appears increasingly severe and more complex in a way correlated between the pixel resolution and the mean particle spacing, e.g., the interplay between damping of power due to low resolution and the aforementioned small scale spurious power. A volume-weighted measurement using the Delaunay tessellation may remedy this problem ~\citep[e.g.,][]{Pueblas:2008uv}, but implementing such method also requires a large number of particles per tessellation pixel, again limiting a pixel resolution. In a future paper, we plan to correct for this window function effect, but in this work, we do not correct for the pixel window function, nor do we subtract shot noise contribution in the case of \DisIter\, and \DisIterss, and we proceed with a warning that our clustering measurement of the reconstructed displacement would be subject to an uncorrected pixel window function effect.

\subsubsection{Iterative Displacement Reconstruction using the reference fields: \DisIterss}\label{subsec:DisIterSS}
In order to mitigate the sparse sampling problem of \DisIter, as discussed in \S~\ref{subsubsec:scarcity} without sacrificing the pixel resolution, we invent and test a surrogate method where we trace the displacement of the reference field $\mathbf{s}_{net}(\bq_0)$ in Eq.~\ref{eq:xdisq}.

There are two options to evaluate such displacement field traced by reference fields. The first is to evaluate it at the initial, uniform positions of the reference particles. We found that this option ultimately reproduce the observed nonlinear field itself after many iterations. Instead, we have to evaluate such displacement field in the final, non-uniform positions of the reference particles:

\begin{align}
	\hat{\mathbf{\chi}}_{\rm net}(\bq_p)
	&= \frac{\sum_i W_{\rm CIC}({\bq_p,\bq_i}) \mathbf{s}_{\rm net}(\bq_i)  }{\sum_i W_{\rm CIC}({\bq_p,\bq_i}) }
	,\label{eq:massdisq}
\end{align}

That is, both for the galaxy particles and reference particles, the reconstructed displacement has to be evaluated in its estimated Lagrangian/original position, just like the true displacement field~\footnote{The true displacement field also is highly correlated with the linear field if it is evaluated in the Lagrangian position, but returns poorly-correlated field if it is evaluated in the final Eulerian position.}, and it is not important that the tracers of the displacement field are uniform or nonuniform.
We call this method as `\DisIterss'.

In the limit of the near perfect reconstruction, $\hat{\mathbf{\chi}}_{\rm net}(\bx_n)$ or $\hat{\mathbf{\chi}}_{\rm net}(\bq_n)$ would be close to the true displacement field and the final positions of particles would be Lagrangian. \citet{Ota:2021} shows that in reality, the reconstructed displacement field does not perfectly converge to the true displacement field. As we will show, the displaced particles at the last step are approximately uniform on large scales, but the uniformity decreases on small scales. .

\subsubsection{Dealing with halo/galaxy bias}\label{subsec:methodbias}
For the biased tracers, after the first reconstruction, the displaced galaxies have moved to their estimated Lagrangian positions. As they are still biased tracers in the Lagrangian positions, however, clustering of the displaced tracers is subject to the Lagrangian bias $b-1$ on large scales:
\begin{eqnarray}
\delta_d &\sim & (b-S(k))\delta_L + ...\\
\delta_s &\sim & S(k)\delta_L+....\label{eq:biasdelta}
\end{eqnarray}
Therefore an iteration based on $\delta_d$ will introduce a substantial, additional large scale displacement to the particles that are already near their Lagrangian positions. We adopt several options in dealing with bias in iterations.

\begin{enumerate}
\item 
The simplest extension of \StdIter\ could be rescaling all steps of $\delta_d$ with $1/b$ just like the first step ($Eq.,~\ref{eq:recisoq}$) and follow the same procedure as \StdIter\ for the matter field. We also tried rescaling all higher iterations of $\delta_d$ with $1/b^2$, to further reduce the magnitude of the large-scale displacement field after the first iteration:
\begin{eqnarray}
&&\skdis{n} = - \frac{i\bk}{k^2}\frac{1}{b^2} {\delta}_{d}^{(n)}(\bk)S^{(n)}(\bk)\mbox{ for  $n>1$.} \label{eq:stditerbias}
\end{eqnarray}
We find that the former (with the $1/b$ rescaling for higher iterations) barely improves
reconstruction (in terms of propagator at $k=0.2\ihMpc$) beyond the single step standard reconstruction along the transverse direction. Moreover, after the second reconstruction ($n > 1$), the performance quickly becomes worse than the standard reconstruction. The latter option (with the $1/b^2$ rescaling for higher iterations), on the other hand, improves until $n=3$ and slowly diverges for higher iterations.~\footnote{In fact, $1/b^2$ does not appear to be special; a factor greater than $b$ that better suppresses the incorrect estimation of the large scale displacement, such as $1/(1.5b)$, gives a similarly optimal performance.} We choose the latter as our \StdIter\ convention for the biased field. Note that this rescaling reduces, but does not completely fix the problem of non-vanishing large scale displacement field in $\skdis{n}$ after the first reconstruction. But as we mentioned, despite the incorrect estimation of the large
scale flow in the second and higher reconstruction steps, the combination of $\delta_d$ and $\delta_s$ still returns an improved propagator with a few iterative steps. The fiducial bias here was estimated from the real-space cross power spectrum between the biased and the matter fields in \citet{Ding2018}. In real data, we may not be able to estimate a precise large-scale bias. But the effect of an incorrect bias within 10\% has been shown small~\citep{Mehta:2011,Sherwin:2019}. 

\item The non-vanishing $\delta_d$ on large scale for the biased tracers (Eq.~\ref{eq:biasdelta}) is much more detrimental for \DisIter\ as we are trying to iteratively recover the displacement field itself in this method. To mitigate this issue, we have to remove the Lagrangian bias. We therefore {\it\bf debias} the observed field before starting the reconstruction; i.e., a debiased density field is assigned to each mesh based on the measured density of the galaxies in that mesh:

\begin{eqnarray}
\frac{\rho_{\rm debiased}}{\rhobar_{\rm debiased}}=\frac{ \frac{\rho_{\rm b}}{\rhobar_{\rm b}} -1}{b}  +1,
\end{eqnarray}
where $\frac{\rho_{\rm b}}{\rhobar_{\rm b}}-1$ is the measured over density field of the galaxies, $\delta^s_{\rm NL}$. We then displace the meshes of this new density field in the subsequent iterations without the need for any further bias correction. It is equivalent to assigning a particle in the middle of each mesh with this debiased density as its weight and displace them.
In the sense that we trace the displacement of reference particles/meshes, but with a debiasing weight, this treatment can be considered as a hybrid between \DisIter\ and \DisIterss. Using {\it debiasing} also reduces the aforementioned effect of sparsity.
For comparison, we also test {\it debiasing} with \StdIter.
\end{enumerate}

We summarize all these methods in Tab.~\ref{tab:method}.

\begin{figure*}
\centering
\includegraphics[width=1.0\linewidth]{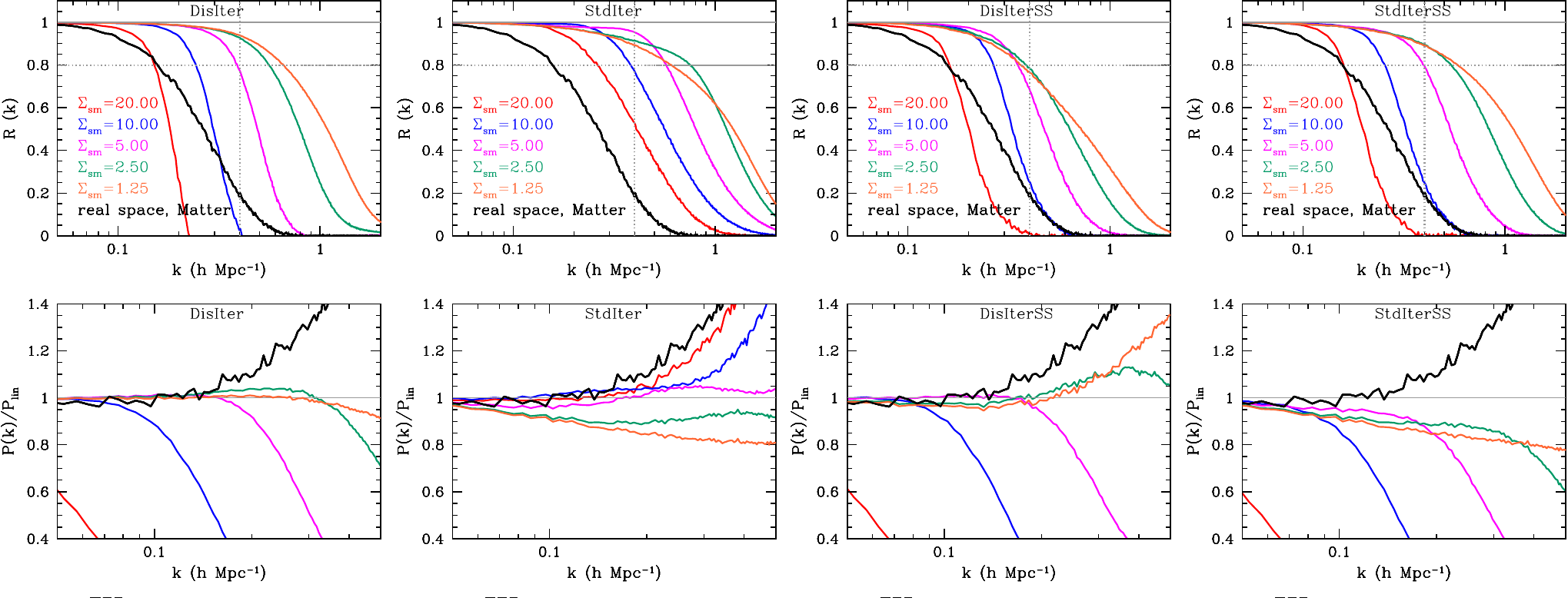}
\caption{Comparison of different iterative schemes for matter in real space. Shot noise is negligible. Top: cross-correlation coefficient. Bottom: power spectrum divided by the linear $P(k)$. Note the different $k$ ranges between the top and the bottom panels. The black lines show the pre-reconstruction measurements and the colored lines show the post-reconstruction fields after different number of iterations. From the left to right, we are showing \DisIter, \StdIter, \DisIterss, and \StdIterss, as summarized in Table~\ref{tab:method}.
The red line in the \StdIter\ case corresponds to the standard, single step reconstruction with $\Sigsm=20\hMpc$.
As reference lines to guide our eyes, the solid grey line marks the unity and the dotted grey lines in the top panels mark $0.8$ at $k=0.2\ihMpc$. 
The iteration was conducted from the initial smoothing scale of $20\hMpc$ while decreasing by $\sqrt{2}$ until it reaches the 9-th reconstruction and the final smoothing scale of $1.25\hMpc$. Here, we are showing only the 1st (red), 3rd (blue), 5th (magenta), 7th (green), and 9th (orange) steps. }
\label{fig:DM500rkreal}
\end{figure*}
\subsection{Estimators for comparison}
In the following section, we will compare the two main and their surrogate iterative reconstruction methods with the optimal case of the standard reconstruction. We use three estimators to evaluate the performance of reconstruction. First, we use the propagator, which is the cross-correlation between the initial density field and the final or reconstructed density field, normalized with the power spectrum of the initial density field:

\begin{eqnarray}
C(\bk)\equiv\frac{<\delta_i(\bk)\delta_z^*(\bk)>}{P_{\rm i}(k)},
\end{eqnarray}
where $\delta_z$ is either the observed late time density field or reconstructed density field and $\delta_i$ and $P_{\rm i}$ are the initial linear density field and the corresponding power spectrum, respectively, after they are scaled with the growth factor. This estimator measures the damping of the BAO, i.e., approximately the shift term $P_{13}/P_{\rm lin}$ contribution.

The second estimator is the cross-correlation coefficient:
\begin{eqnarray}
R(\bk)\equiv \frac{<\delta_i(\bk)\delta^*_z(\bk)>}{\sqrt{P_{i}(k)P_{z}(k)}},
\end{eqnarray}
where $P_z$ is the observed power spectrum without shot noise subtraction. This cross-correlation coefficient can be considered as the signal-to-noise weighted propagator in a Gaussian limit. In addition, this estimator approximately measures the 1-loop mode-coupling contribution $P_{22}$, as pointed out by \cite{Ota:2021}:

\begin{eqnarray}
	R(k) & \sim & 1 -\frac{1}{2}\frac{P_{22}}{P_{\rm lin}}
\end{eqnarray}

Finally, we are also checking the power spectrum and correlation function of each case to inspect the net broadband shape of the resulting clustering as well as the BAO feature.

In Fourier space, we inspect the transverse modes by looking at the modes with $0.1 \leq \mu < 0.2$ and the line of sight modes by $0.9 \leq \mu < 1$, where $\mu=\hat{\bk}\cdot{\hat{\mathbf z}}$.
Inspecting the modes in separate $\mu$ bins increases the sample variance. Whenever we separate the modes, we therefore reduce the noise by applying Savitzky-Golay smoothing filtering \citep{SaGo} for a better visualization. 
In some plots, we will estimate the characteristic BAO damping scale $\Sigma_{\rm nl,eff}$ before and after reconstruction using $C(k=0.2\ihMpc)$ in each of the two $\mu$ bins:
\begin{equation}
C(k_{0.2})=(1+\beta\mu^2) \exp\left[-\frac{k_{0.2}^2\Sigma_{\rm nl,eff}^2}{4}\right],\label{eq:CkG}
\end{equation}
such that $\Sigma_{\rm nl,eff}$ from $\mu=0.1-0.2$ ($\mu=0.9-1$) approximately measures the transverse (the line-of-sight) BAO damping scale when defined at $k=0.2\ihMpc$. This should not be taken as an accurate measure of the performance, especially when the measured damping deviates from a Gaussian damping  (e.g., Eq. \ref{eq:CkG}) or when there is a noisy feature at $k=0.2\ihMpc$. Therefore, we also inspect $C(k)$ and $R(k)$ at $k=0.4\ihMpc$.

 All spherically averaged multipoles (e.g., for the BAO feature inspection) presented are not smoothed using this filtering. 
In the configuration space, we present the Legendre multipoles  $\xi_\ell$. 
Taking advantage of the periodic boundary condition, we simply conduct a 3-dimensional Fourier Transformation to derive $\xi(\vec{r})$ from the measured $P(\vec{k})$ and take the Legendre decomposition.

\subsection{Simulations}
We use three sets of simulations in this paper, which are listed in Table~\ref{tab:sim}. We utilize multiple sets of simulations, instead of a single consistent set,  due to our limited computational resources, to inspect different aspects of the reconstructed clustering. They all assume a flat $\LCDM$ cosmology based on \citet{2016A&A...594A..13P} with $\Om = 0.3075$, $\Obhh=0.0223$, $h=0.6774$, and $\sigma_8=0.8159$. 

\begin{itemize}
\item Full N-body simulation using the MP-Gadget code~\citep{FengMPgadget} with the box volume of $500\hMpc$. We use the average of five simulations. The simulation evolves $1536^3$ particles from $z=99$ by computing forces in a grid of $1536^3$ and, to reduce the data storage and the computational time/memory for the analysis, we subsample 4\% of the output particles at $z=0.6$~\footnote{C.f. \cite{Schmittfull:2017} at $z=0$}.  This set is called `L500' in this paper. We use a grid of $512^3$ to Fourier-transform and reconstruct this nonlinear field. With 4\% subsampling, there is approximately one particle per grid of $512^3$. This set of simulations was used to reach  the highest particle and mesh resolution  and the lowest shot noise to compare with the result of \citet{Schmittfull:2017}.  We also test 0.15\% of the output particles to test the effect of sparsity on reconstruction and we call this sample `subL500'.
\item Full N-body simulation using the MP-Gadget code with the box volume of $1500\hMpc$. We call this simulation `L1500'. This simulation is mainly used for the BAO feature inspection, as the volume of L500 is believed to be small for the robust BAO feature. This simulation also evolves $1536^3$ particles from $z=99$ by computing forces in a grid of $1536^3$ and we subsample 4\% of the output particles at $z=0.6$. We use a grid of $1024^3$ to reconstruct and therefore only 13.5\% of all the grids/meshes contain a mass tracer on average. We use a pair of simulations that match in terms of phase and the broad-band shape, one with the BAO feature in the initial field and one without the BAO feature~\citep{Prada2016,Ding2018}. The cosmic variance as well as any spurious effect on the broadband shape due to sparsity (e.g., for \DisIter) will largely cancel out. 
\item FastPM simulation. We use two halo catalogs of the FastPM simulations~\citep{FengFastPM} used in \citet{Ding2018} and \citet{Schmittfull:2017}. The simulation uses a box of $1380\hMpc$ and evolves $2048^3$ particles through 40 time steps linearly spaced between $a=0.1$ and $a=1$ by computing forces on a $4096^3$ particle-mesh grid. FastPM is a quasi-\Nbody\ simulation which models the evolution of dark matter non-perturbatively by employing a Particle-Mesh solver with a finite number of time steps, when compared to the full \Nbody\ such as MP-Gadget~\citep{FengFastPM}.  \citet{Ding2018} shows that the FastPM simulations we use are cross-correlated with the full \Nbody\ simulation at the level better than 96\% at $k < 0.3\ihMpc$. Therefore our results of the biased cases could be well subject to this level of error for $k>0.3\ihMpc$.  We use output halo catalogs with $b= 1.48$ and 1.88 at $z=1$. The grid used for reconstruction is $512^3$.
\end{itemize}

In all cases of the iterative reconstruction, the iteration was conducted with the initial smoothing scale of $20\ihMpc$ and the smoothing scale decreased by $\sqrt{2}$ until it reaches the 9-th reconstruction and the final smoothing scale of $1.25\hMpc$. As a caveat, \citet{Schmittfull:2017} used the initial smoothing scale of $14\hMpc$ (equivalent to $10\hMpc$ in their definition). All simulations are used for the real and the redshift-space comparisons. For the matter fields (i.e., other than FastPM), the reconstruction is performed in the presence of a substantial finger-of-God effect, while the FastPM halo catalogs that select halo centers suffer a relatively smaller level of finger of God.

\section{Results}\label{sec:Results}
\begin{figure*}
\centering
\includegraphics[width=1.0\linewidth]{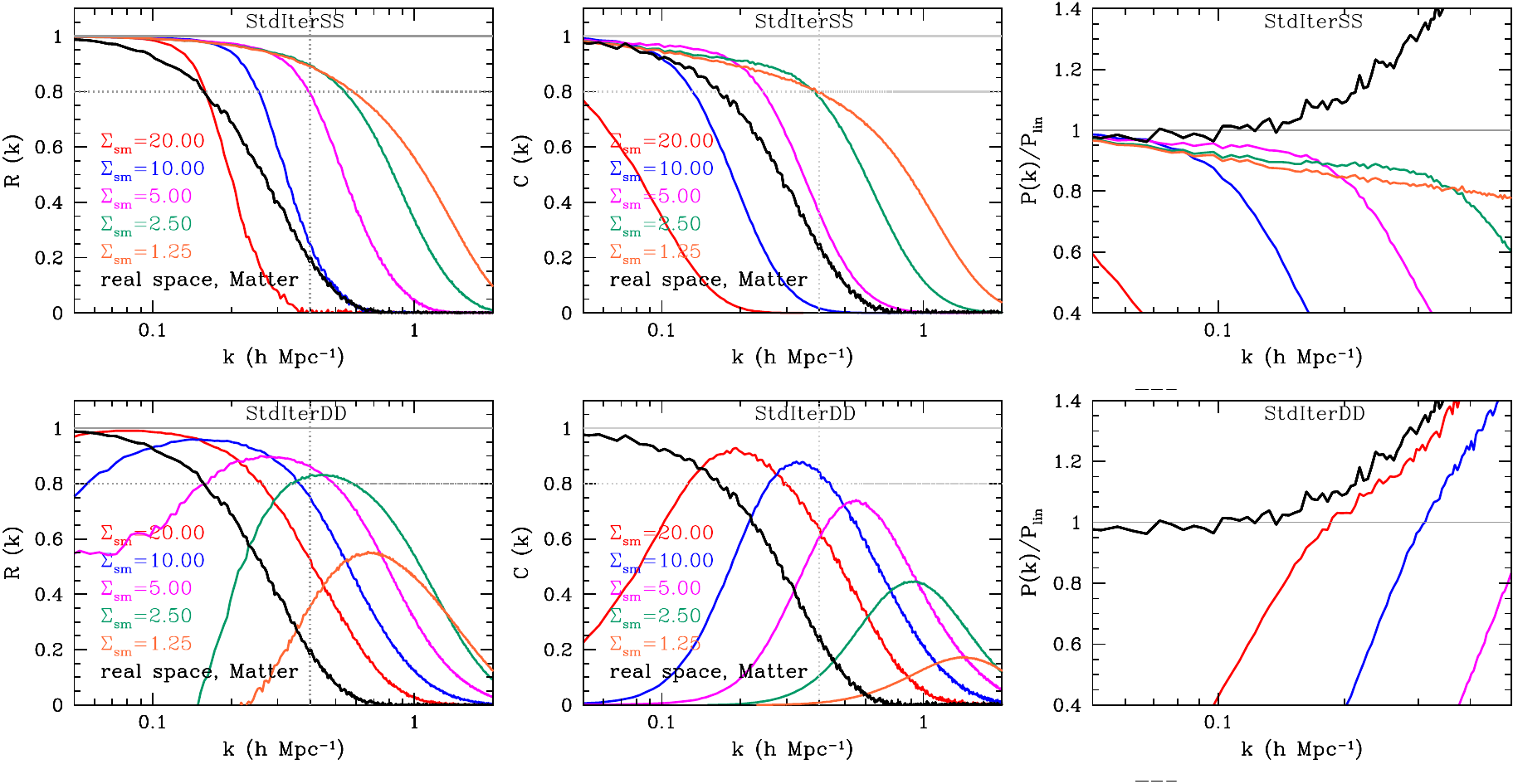}
\caption{The components of \StdIter\ from Fig.~\ref{fig:DM500rkreal} to show how the mass and the reference field evolves for all cases of iterative reconstructions. The top panels show the properties of the displaced reference/random fluctuation in each step (i.e. \StdIterss) and the bottom panels show the displaced galaxy fields (i.e. the DD component). In each step of \DisIter\ and \StdIter\, the displacement of the next iteration is constructed based on the DD component shown in the bottom panels. }
\label{fig:DM500realssdd}
\end{figure*}

\subsection{Iterative reconstruction of the matter field}
\subsubsection{Real space}
We first apply the different iterative reconstruction schemes introduced in the previous section to the real-space matter field to check the optimal limit of each method. In Figure \ref{fig:DM500rkreal}, we present the cross-correlation coefficient $R(k)$ (top) and the power spectrum (bottom) at $z=0.6$  using L500 that has negligible shot noise (Table \ref{tab:sim}). While our performance evaluation will be mainly qualitative, we add the guiding lines that mark $R(k)=0.8$  at $k=0.4\ihMpc$ and quantitatively compare the performance at this wavenumber whenever possible.

The left panel shows the iterative displacement reconstruction, \DisIter, can recover  $R(k)\sim 0.9$ at $k\sim 0.5\ihMpc$ at the final smoothing scale, as observed by \cite{Schmittfull:2017}. The constraint wave number $k_C$ \footnote{The constraints wavenumber $k_C$ is defined such that the number of modes smaller than $k_C$ equals the number of constraints: $k_C\equiv 0.4\left(\frac{N}{4}\right)^{1/3}\left(\frac{\nbar}{10^{-3}{\rm Mpc}^{-3}}\right)^{1/3}$ where $N=3$ representing 3 positions for each galaxy.} suggested by \cite{McQuinn:2020} for this setup is $3.59\ihMpc$, in agreement with the efficient reconstruction  we observe at high $k$. Note that \DisIter\ is not efficient with large smoothing scales in the first few iterations, compared to the \StdIter\ (second column), but then it quickly improves with increasing iterations, being as efficient as or perhaps better than the other cases at the final step, judging based on $k\gtrsim 0.4\ihMpc$.

The second column shows the standard iterative reconstruction (\S~\ref{subsec:StdIter}). \StdIter\ quickly reaches its best performance in a few iterations (we identify the magenta line for $\Sigsm=5\hMpc$, i.e., in the fifth reconstruction, as the best performance in this case), but then starts degrading on a large scale as the iteration continues. We could potentially remedy this degradation by setting a minimum smoothing scale.  Note that the red line of \StdIter, i.e., the first step of \StdIter\ is the single step standard reconstruction with $\Sigsm=20\hMpc$, returning $R(0.4)=0.53$, when the iterative reconstruction with $n=5$ (magenta) shows $R(0.4)=0.95$. 

The third and fourth panels show the two surrogate methods, using only the displaced reference/random particles; i.e., \DisIterss\ (\S~\ref{subsec:DisIterSS}) tracing the displacement field of the reference particles and \StdIterss\ (\S~\ref{subsec:StdIterSS}) being the reconstructed density field of the reference/random particle of \StdIter. These alternatives, in particular \DisIterss\, are slightly less efficient for $k \gtrsim 0.4 \ihMpc$ than their primary counterparts, but they still show $R(0.4) > 0.8$ at $k < 0.4\ihMpc$.  It is because, after many iterations, the information has been effectively transferred from the displaced mass field to the displaced random field. An advantage of \DisIterss\ compared to \DisIter, is that it does not suffer the empty pixel problem in deriving the displacement field.  As an advantage of \StdIterss, it is a byproduct of \StdIter, it requires only one field in the final step, and has a more consistent convergence behavior with decreasing smoothing scale during iteration.

Fig. ~\ref{fig:DM500realssdd} shows the individual component of \StdIter: the displaced reference/random component is equivalent to \StdIterss\ in the top panel and the displaced mass field, which we call `StdIterDD' or 'DD' for convenience, is in the bottom panel. One can see that, in $C(k)$ (middle column), the initial information in the displaced mass field decreases and shifts to a smaller scale with increasing iterations (and therefore, a decreasing smoothing scale). Interestingly, the cross-correlation $R(k)$ of the displaced mass (bottom left) during the first few iterations remains near unity at small $k$, implying the decrease in the linear information is proportional to the decrease in power so that the signal to noise does not decrease under the Gaussian approximation. Moreover, $R(k)$ of the DD component looks even better than $R(k)$ of the SS component for the first three reconstructions. In Fig.~\ref{fig:xiiter}, we will show that, indeed, most of the BAO feature seems to reside in the displaced mass field after the first reconstruction when a large enough smoothing scale, e.g., $20\hMpc$, is used.

Fig.~\ref{fig:DM500rkreal} and \ref{fig:DM500realssdd} also show the power spectrum of all cases. We find that \DisIter\ returns the most stable agreement with the linear $P(k)$ for $k < 0.3\ihMpc$, while \StdIter\ shows the most complex behaviour. As a caveat, if \DisIter\ recovered the true nonlinear displacement field, its power spectrum should deviate from the linear $P(k)$~\citep{Baldauf:2015dis}. The fact that \DisIter\ converges close to the linear $P(k)$ implies that we are not quite recovering the nonlinear displacement field, even though we are pulling out most of the linear information in the nonlinear displacement field. The very small, but nonzero power of the DD component at the 9-th reconstruction (i.e., the orange line in the bottom middle panel of Fig.~\ref{fig:DM500realssdd}) implies that we are not perfectly recovering the uniform Lagrangian distribution on small scales. For more discussion on this aspect, we refer the readers to \citet{Ota:2021}.

In summary, we find that all the iterative methods we are testing are performing comparably at the negligible shot noise limit when redshift-space distortions and galaxy/halo bias are not included. Among these options, \DisIter\ shows the best behaviour in terms of  the convergence in $R(k)$ and $P(k)$ to the linear field.  The single-step standard reconstruction performs significantly worse than the later-stage iterative reconstruction when the starting smoothing scale of the iteration is the same as the single-step smoothing scale, confirming the result in \citet{Schmittfull:2017}. This is expected as the effective smoothing scale used is very different between the two cases. The difference we find could be somewhat more severe than \citet{Schmittfull:2017}, as they used an initial smoothing scale of $14\ihMpc$ that is more optimal for the standard reconstruction. A more fair comparison, however, would be a comparison between the best performance iterative reconstruction and the best performance single step reconstruction at any smoothing scale. In the next section, we will make such a comparison after including redshift-space distortions and shot noise.

\begin{figure*},
\centering
\includegraphics[width=1.0\linewidth]{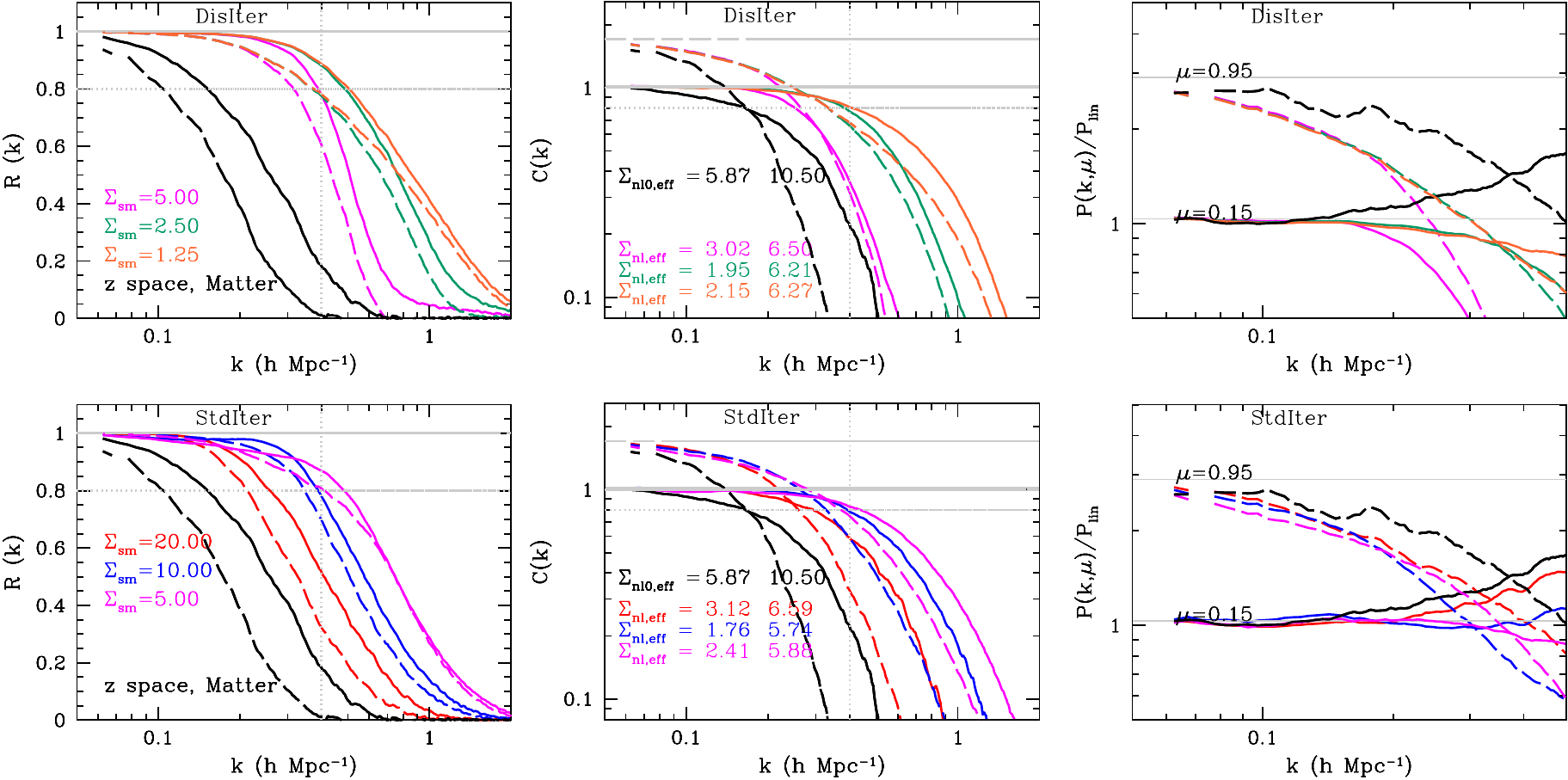}
\caption{Effect of including redshift-space distortions in the iterative reconstruction. We focus on \StdIter\ and \DisIter.  The  iterative steps were selected based on Fig.~\ref{fig:DM500rkreal} for each method: 5th, 7th, 9th for \DisIter\ and 1st, 3rd, 5th for \StdIter. The same color scheme for each iterative step as in Fig~\ref{fig:DM500rkreal}. The top panel shows \DisIter\ and the bottom panel shows \StdIter. The solid lines are for the modes with $\mu=0.1-0.2$ and the dashed lines are for the modes with $\mu=0.9-1.0$. Note that we have changed the y-axis scaling of $C(k)$ and $P(k)$ to a logarithmic scale. To ease the comparison in the logarithmic scaling, we quote the values of $\Sigma_{\rm nl,eff}$ (in $\hMpc$) for each $\mu$ bin (E.q.~\ref{eq:CkG}) in the middle panels to approximately represent the performance of $C(k)$ at $k=0.2\ihMpc$. For example, the two values of $\Sigma_{\rm nl,eff}$, 3.19 and 8.13, in magenta in the top middle panel represent the damping scales in $\hMpc$ for $\mu=0.1-0.2$ and $\mu=0.9-1$, respectively, after the 5th reconstruction with $\Sigsm=5\hMpc$. }
\label{fig:DM500rkred}
\end{figure*}

\begin{figure}
\centering
\includegraphics[width=1\linewidth]{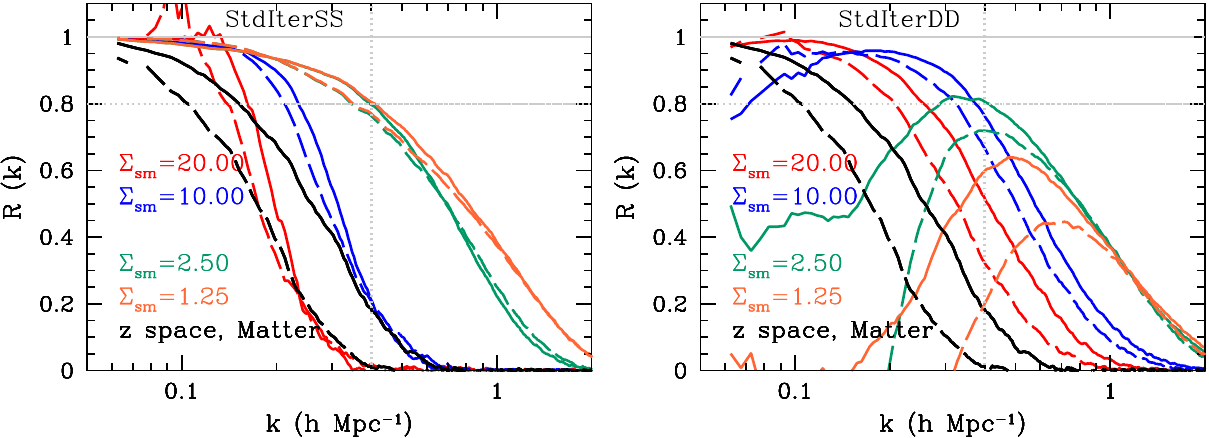}
\caption{The SS and DD component of \StdIter\ in redshift space at three snapshots of iteration to show how the mass and the reference field evolves for iterative reconstructions. The solid lines are for the modes with $\mu=0.1-0.2$ and the dashed lines are for the modes with $\mu=0.9-1.0$. }
\label{fig:DM500redcomp}
\end{figure}

\subsubsection{Redshift-space distortions}
We then move to the redshift space. We select three iterative snap shots for each method that we subjectively consider representing the steps to its optimal performance in real space. We also focus on the two primary iterative methods, \DisIter\ and \StdIter.  Fig.~\ref{fig:DM500rkred} shows the three estimators in the redshift space along the transverse direction $\mu=0.1-0.2$ (solid line) and $\mu=0.9-1.0$ (dashed line). Note that we change the y-axis scaling of $C(k)$ and $P(k)$ to a logarithmic scale hereafter, as the range of the scale to cover increased. As a result, these plots are less sensitive to a small offset when $C(k)$ and $R(k)$ reach near unity. To ease the comparison , we quote the values of $\Sigma_{\rm nl,eff}$ (E.q.~\ref{eq:CkG}) in the middle panel to approximately represent the performance of $C(k)$ at $k=0.2\ihMpc$, but these estimates do not necessarily represent the performance on much smaller scales. When compared to Fig.~\ref{fig:DM500rkreal}, focusing on $C(k)$ and $R(k)$, even the nearly transverse Fourier modes are degraded in terms of its cross-correlation with the initial field when the reconstruction is done based on the density field measured in the redshift space; for \DisIter, $R(k)=0.8$ at $k=0.65\ihMpc$ in real space, but at $k=0.5\ihMpc$ in redshift space. For the nearly line-of-sight modes ($\mu=0.9-1.0$), the performance is much more degraded: $R(k=0.37\ihMpc)=0.8$.

In terms of power spectrum, we find that, for \DisIter\, the iteration tends to bring up the damped power, while \StdIter\ tends to damp small-scale power more for the first few iterations along the line of sight, as if it makes the finger-of-God effect more severe. With increasing iterations, both methods seem to converge with respect to each other in terms of $P(k)$, which could be a mere coincidence as a further iteration with \StdIter\ beyond the magenta line makes the convergence worse.

Fig.~\ref{fig:DM500redcomp} shows that in terms of cross-correlation $R(k)$, after the first reconstruction, again, the DD field alone contains most of the signal to noise for the recovered initial information given the smoothing scale, which is similar to Fig.~\ref{fig:DM500realssdd}. Along the line of sight, we find that the trend persists even for the second and the third reconstructions. This trend of the DD component including most of the information in the beginning disappears when the smoothing scale decreases, either during iteration or by decreasing the initial smoothing scale.

\subsubsection{Iterative reconstruction in comparison to the standard reconstruction}
\begin{figure*}
\centering
\includegraphics[width=0.9\linewidth]{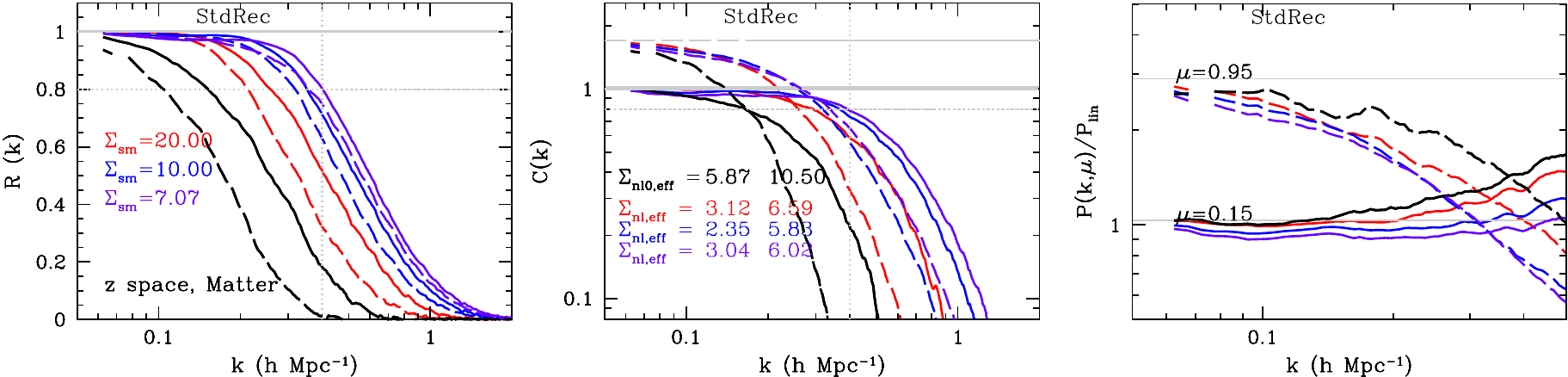}
\caption{The single step standard reconstruction. Here we focus on $\Sigsm=20$, $10$, and $7\hMpc$, which are the best cases of the standard reconstruction we identified for our setup. The solid lines are for the modes with $\mu=0.1-0.2$ and the dashed lines are for the modes with $\mu=0.9-1.0$. When compared to Fig.~\ref{fig:DM500rkred}, the single step standard reconstruction with $\Sigsm=7\hMpc$ appears nearly as good as the \StdIter, while beginning to show a large scale deviation, which is more obvious in $P(k)$. }
\label{fig:DMrsdSingle}
\end{figure*}

So far, we have observed the iterative reconstructions improving the reconstruction gradually in each step while the smoothing scale is being updated towards a smaller scale. In this section, we compare the iterative reconstruction and the single-step standard reconstruction.
Fig.~\ref{fig:DMrsdSingle} shows a few examples of the standard reconstruction with different smoothing scales. $\Sigsm=20\hMpc$ corresponds to the first step before iteration and performs much worse than the iterative reconstruction. For a more fair comparison, we single out a few optimal cases of the standard reconstruction in terms of the maximum cross-correlation over a broader $k$ range; in our setup, we identify it to be 10 or $7\hMpc$~\footnote{The optimal smoothing scale will depend on the shot noise and the redshift.}, as shown in Fig.~\ref{fig:DM500rkred}.  When comparing Fig.~\ref{fig:DMrsdSingle} and Fig.~\ref{fig:DM500rkred}, it appears that the optimal standard reconstruction performs nearly as well as the optimal iterative reconstruction in terms of $C(k)$ and $R(k)$, giving $R(k)=0.8$ at $k=0.4\ihMpc$ for the transverse modes when using $\Sigsm=7\hMpc$. On a closer look, however, one can notice that the standard reconstruction, using $7\hMpc$ shows  degradation in correlation with the initial field on much larger scales than the iterative methods do. The degradation is about 9\% in $C(k)$ for $k\sim 0.1\ihMpc$ transverse modes, compared to $\sim 2\%$ of \DisIter\ (the 9th step) and $\sim 4\%$ for \StdIter\ (the 5th step).
The large-scale degradation is most obvious in the $P(k)$ plot.  Near $k=0.1\ihMpc$, focusing on the transverse modes, the deviation from the linear theory $P(k)$ is at the level of 13\% for the standard reconstruction using $7\hMpc$, compared to 3-4\% for the optimal \DisIter\ and \StdIter. A degradation of the standard reconstruction with a small smoothing kernel in terms of agreement with the theory BAO fitting model has been observed previously \citep[e.g.,][]{Seo:2016brs} and \citet{Hikage:2017real} reproduced such trend using a PT theory.~\footnote{$5\hMpc$ of the smoothing scale in \citet{Hikage:2017real}  corresponds to our $7\hMpc$.} Our result is consistent with these earlier findings. The offset from the linear $P(k)$ and  $C(k)$  on large scales could potentially affect the goodness of the BAO fitting, e.g., the goodness of the empirical, PT-based BAO damping model for the BAO-only analysis and certainly the full post-reconstruction power spectrum modeling unless we can correctly account for it. 

To summarize, we find that when reconstruction is aggressively performed to exploit as much information as available, taking the iterative step allows us to do so more `stably', i.e., without causing a substantial deviation from the linear theory model on large scales. Such stability can be advantageous in terms of the goodness of fit for the BAO-only analysis as well as the broadband-RSD analysis using the post-reconstructed field. 

\begin{figure*}
\centering
\includegraphics[width=0.9\linewidth]{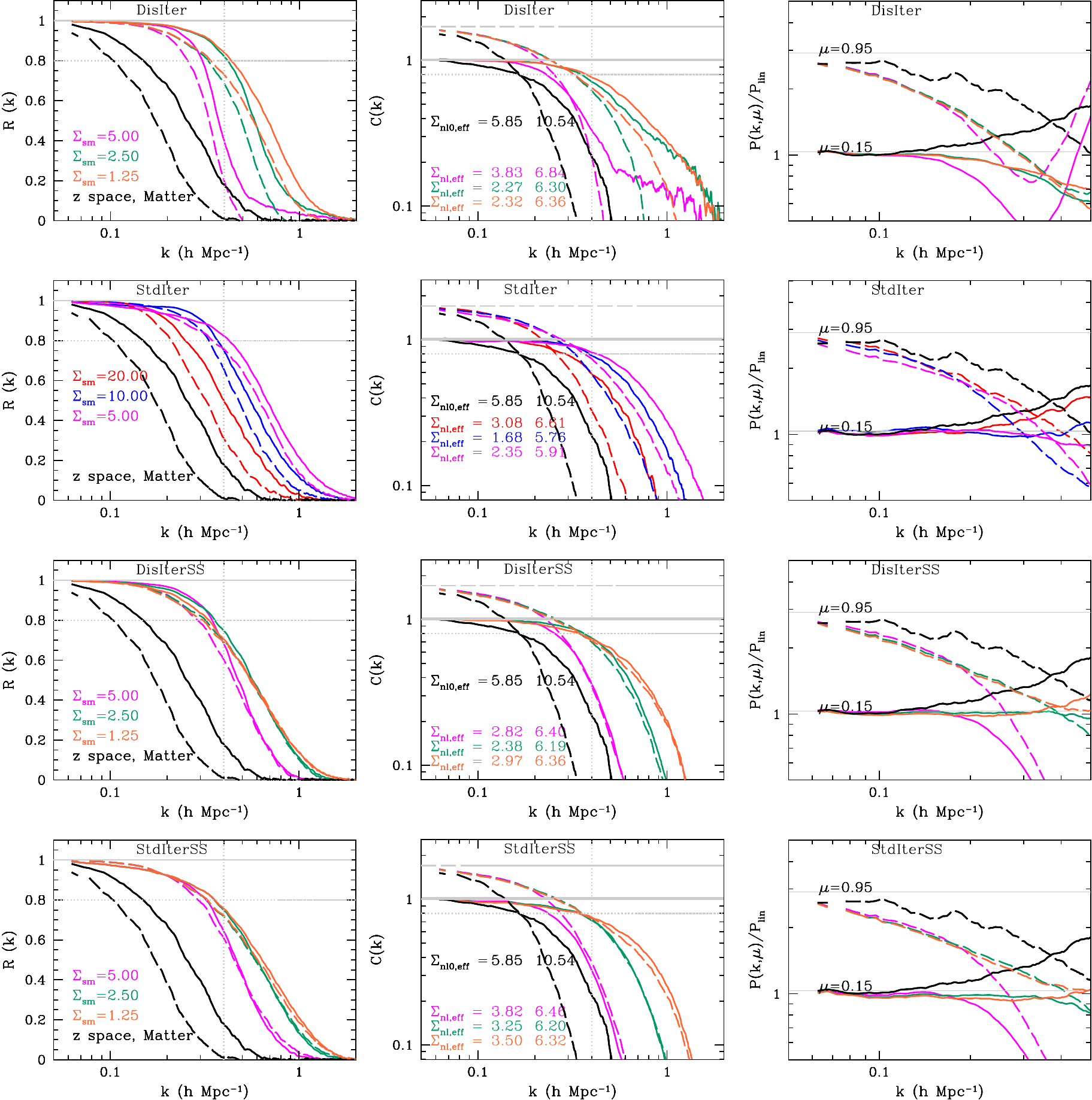}
\caption{The effect of an incremental shot noise. Using $subL500$ with $\bar{n}=0.0442\itrihMpc$. Cross-correlation coefficient, propagators and power spectra of the three iterative schemes for matter in redshift space. The propagator and power spectrum of the \DisIter\ shows that this method suffers an artifact on small scale even at this low level of shot noise.     }
\label{fig:DM500redcompsub}
\end{figure*}

\begin{figure*}
\centering
\includegraphics[width=0.9\linewidth]{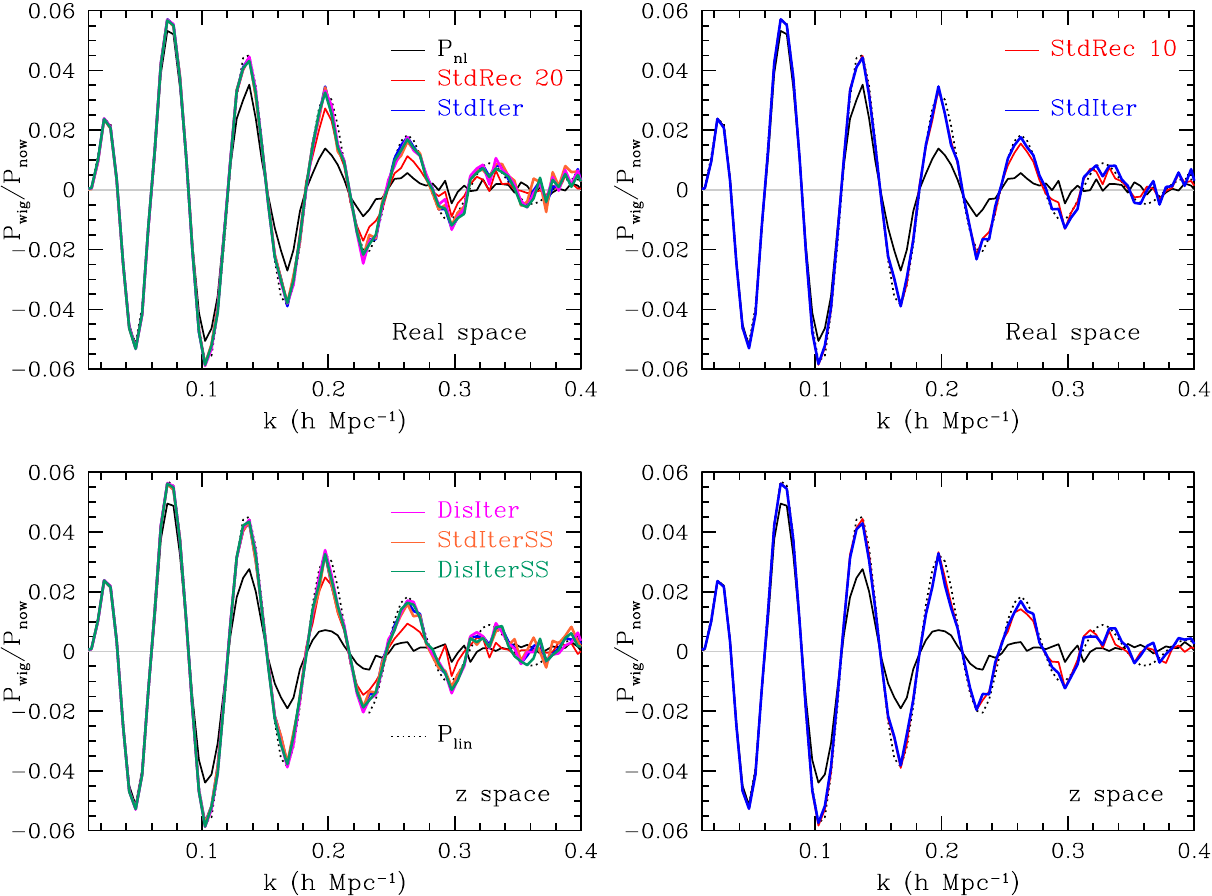}
\caption{The BAO feature in the various iterative methods, in comparison to the standard, single step reconstruction (matter). Top: spherically averaged real space. Bottom: spherically averaged redshift space. The left panels show all four iterative methods with its optimal final smoothing length in comparison to the single step, standard reconstruction using the same initial smoothing scale (red, $20\hMpc$). The dotted line is the input power spectrum which is nearly invisible as overlaid by other lines, the black solid line is the nonlinear power spectrum at $z=1$. All iterative methods started from $\Sigsm=20\hMpc$. Blue: \StdIter\ that ended at $\Sigsm=5\hMpc$ (5th step). Orange: \StdIterss\ with $\Sigsm=1.25\hMpc$ (9th step). Magenta: \DisIter\ (9th step) with $\Sigsm=1.25\hMpc$. Green: \DisIterss\ with $\Sigsm=2.5\hMpc$. The right panels show the comparison between \StdIter\ with $\Sigsm=5\hMpc$ and the optimal single step standard reconstruction with $\Sigsm=10\hMpc$. } 
\label{fig:DMPwPnw}
\end{figure*}

\subsubsection{Shot noise}\label{subsubsec:shotnoise}
We will show the effect of more realistic shot noise and galaxy bias in \S~\ref{sec:bias}, but here we introduce an incremental shot noise and observe how the performance of the iterative reconstruction changes due to this shot noise. 
Fig.~\ref{fig:DM500redcompsub}, in comparison to Fig.~\ref{fig:DM500rkred}, shows the effect of increasing the shot noise from $\nbar=1.18\itrihMpc$ to $0.0442\itrihMpc$ using $subL500$. 
With such a small additional shot noise, most of the pixels (i.e., 96\%) of $subL500$ become empty if we want to reach a mesh resolution of $\sim 1\hMpc$. This problem can be treated properly for density field-based reconstructions such as \StdIter\ and \StdIterss, but \DisIter\ lacks the tracers of the displacement field in the majority of the pixels and this introduces an artifact in $P(k)$ as the high $k$ upturn shown in the top right panel of Fig.~\ref{fig:DM500redcompsub} (as discussed in \S~\ref{subsubsec:scarcity}). \DisIterss\ (third row) does not show such a strong spurious feature, as the displacement field is traced by the reference particles that are prevalent. We therefore adopt a mesh-based hybrid between \DisIter\ and \DisIterss\ for the biased tracers in \S~\ref{sec:bias}.

\begin{figure*}
\centering
\includegraphics[width=1\linewidth]{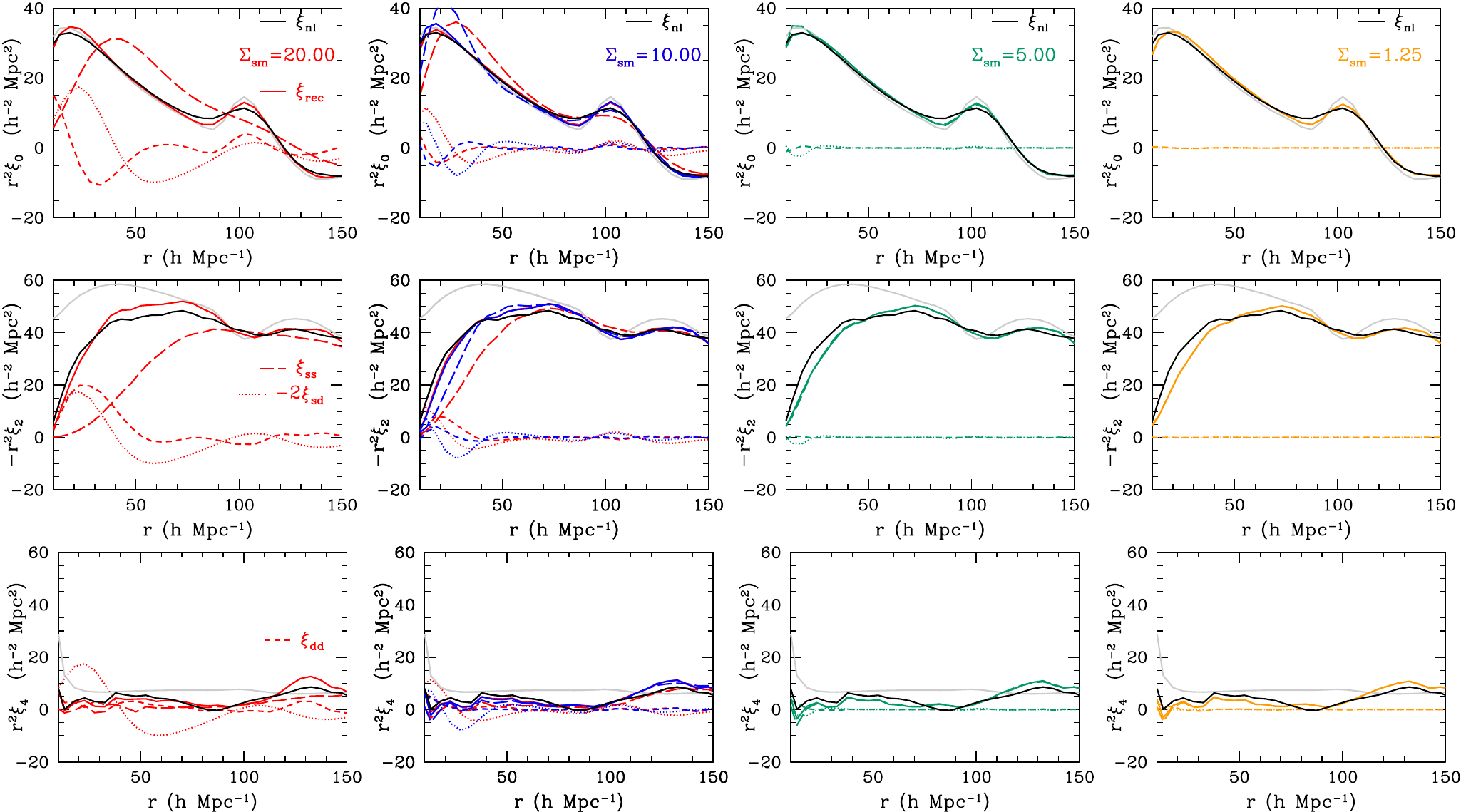}
\caption{The iterative reconstruction in the configuration space for matter in the redshift space. The top, middle, and the bottom rows show $\xi_0$, $\xi_2$, and $\xi_4$ in the redshift space, respectively, before (black line) and after reconstruction (colored). The gray line is the input, linear correlation function with linear RSD prediction.  The red line in the first column is the standard reconstruction using $\Sigsm=20\hMpc$ while blue (2nd column), green (3rd), orange (4th) solid lines show the snapshots of the iterative reconstructions, \StdIter, at $\Sigsm=10\hMpc$, $5\hMpc$, and $1.25\hMpc$, respectively when the reconstruction started from $\Sigsm=20\hMpc$. The long-dashed lines show the $\xi_{ss}$ component, and the short-dashed lines for the $\xi_{dd}$ component and the dotted lines for $-2\xi_{ds}$. In the second column, the overlaid red lines show the single-step standard reconstruction with $\Sigsm=10\hMpc$.  }\label{fig:xiiter}
\end{figure*}

\subsection{The BAO feature}

In the previous section, we identified the advantage of the iterative reconstructions at the small smoothing scale limit in terms of the agreement with the linear theory model on large scales. While the propagator comparison indirectly indicated a comparable reconstructed BAO feature between the iterative and standard reconstruction, we want to directly inspect the resulting BAO feature in each method for the consistency check. 

We use a pair of wiggle and nowiggle $L1500$ simulations to single out the BAO feature in different reconstruction conventions. In Fig~\ref{fig:DMPwPnw}, we compare the BAO feature in various iterative methods, in comparison to the standard reconstruction.  The top and bottom panels show the spherically averaged real space and redshift space, respectively. The left panels show all four iterative methods with its optimal final smoothing length we chose in comparison to the single-step standard reconstruction using the same initial smoothing scale (red, $20\hMpc$). Dotted line is the input power spectrum, the black solid line is the nonlinear power spectrum. All iterative methods start from $\Sigsm=20\hMpc$. Blue, orange, magenta, and green show the optimal iterative reconstruction steps we identified in the previous sections. The right panels again show \StdIter\ with $\Sigsm=5\hMpc$, but this time in comparison to the optimal standard reconstruction with $\Sigsm=10\hMpc$. This figure shows that in terms of the BAO feature alone, the optimal standard reconstruction (blue line) indeed appears to contain as much of the BAO signal as the iterative methods. The actual signal to noise of the BAO measurement would also depend on the covariance structure of the post-reconstructed field, which we will investigate in a future paper. 

Note that any effect on the broadband due to using a small smoothing scale has been canceled out in this figure. 
Given the near equally well-constructed BAO signal for different methods in its own optimal setup, choosing the iterative reconstruction versus the standard reconstruction would mainly depend on the consideration of the broadband modeling. We could aim at a method that produces a power spectrum that is in a better agreement with the linear theory model and/or could aim at a method that gives an easier PT model construction. The iterative step certainly makes the PT model construction more challenging~\citep{Ota:2021}, but the simulation result shows a better convergence to the linear theory model on large scales. On the other hand, the PT model performs promisingly well for the standard reconstruction, but becomes increasingly worse at the smoothing scale that can give the BAO feature comparable to the iterative reconstruction. For example, the SPT model in \citet{Hikage:2017real} works well for $\Sigsm \sim 10\hMpc$ (our $14\hMpc$) at $z=1$ and the Zeldovich approximation in \citet{Chen:2019rec} performs well for $\Sigsm=15\hMpc$ (our $21\hMpc$)  at $z=0$ for describing the full power spectrum over $k<0.2-0.4\ihMpc$, but in both cases, the agreement worsens with a smaller smoothing scale. Weighing up these pros and cons of the iterative steps will require further investigation.

\subsection{Configuration space picture}
Using $L1500$, we observe how these iterative reconstructions appear in the configuration space. We focus on  \StdIter, and compare its components with the standard reconstruction. From the previous sections (e.g., Fig.~\ref{fig:DM500rkreal}), \StdIterss\ (the SS component of \StdIter) was found to be qualitatively similar to \DisIterss\ and  \DisIter,  so showing \StdIterss\ is sufficient for understanding the configuration space picture of the other iterative methods. 

Fig.~\ref{fig:xiiter} shows the multipoles of pre- and post-reconstructed correlation functions in the redshift space. The top, middle, and bottom rows show $\xi_0$, $\xi_2$, and $\xi_4$ in the redshift space, respectively, before (black line) and after reconstruction (colored). The gray line is the input linear correlation function for linear RSD prediction.  The red (1st column), blue (2nd), green (3rd), orange (4th) solid lines show the snapshots of \StdIter. The long-dashed lines show the $\delta_{ss}$ component (\StdIterss), and the short-dashed lines show the $\delta_{dd}$ component. In the second column, the overlaid red lines show the standard reconstruction with $\Sigsm=10\hMpc$.  

In the case of the standard reconstruction with $\Sigsm=20\hMpc$ (the red lines in the first column), the DD component $\xi_{dd}$ (and therefore $\xi_{ds}$) contains the small scale clustering information for $r<50\hMpc$ as well as most of the BAO-like peak. On the other hand, $\xi_{ss}$ provides a smooth curve. With a smaller smoothing scale (red line for $\Sigsm=10\hMpc$ in the second column), $\xi_{ss}$ contains more of the BAO-like feature.  

In the iterative reconstruction, as we noticed in the previous sections, the BAO-like information in the DD component transfers from $\delta_{dd}$ to $\delta_{ss}$, so that both $\xi_{dd}$ and $\xi_{ds}$ vanish at its limit (see the very right column). Again,  we  can view the iterative reconstruction as an operation to transfer and merge information from two density fields into one field.

\begin{figure*}
\centering
\includegraphics[width=0.9\linewidth]{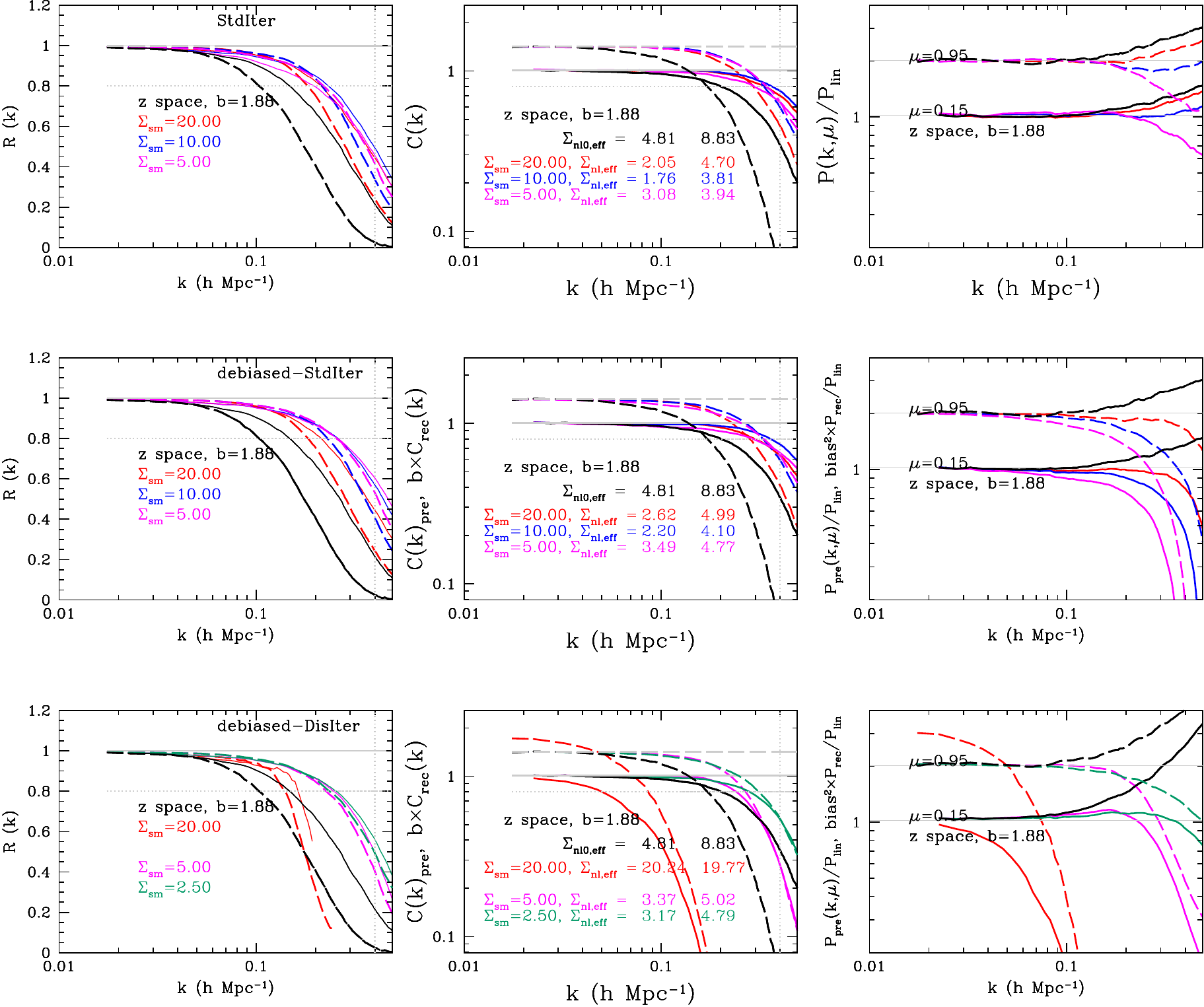}
\caption{Cross-correlation coefficient, propagator, and power spectrum of the iterative reconstruction on the biased sample at $z=1$ with $b=1.88$. Top panel: \StdIter. Middle: \StdIter\ after debiasing. Bottom: \DisIter\ after debiasing. In order to compare those cases with pre-reconstruction, we rescale the debiased cases with the bias factor (i.e., the colored  $C(k)$ and $P(k)$ lines of the middle and the bottom panels). For the debiased-\StdIter $P(k)$, the shot noise is subtracted after scaling with $b^2$. For the debiased-\DisIter, we do not subtract any shot noise from the reconstructed $P(k)$ in the bottom right panel. In the same plot, the pre-reconstruction $P(k)$ is plotted without shot noise subtraction.}\label{fig:biased}
\end{figure*}

\subsection{Iterative reconstruction of the biased field}\label{sec:bias}

The biased field reconstruction will require an additional implementation for dealing with the biased tracers as well as more severe sparsity. We change our mock catalog to FastPM-based catalog at $z=1.0$ with $b = 1.88$ and  $\nbar=0.0012\itrihMpc$~\citep{Ding2018}, i.e., going slightly higher in redshift, accounting for the future galaxy surveys shifting toward higher redshift tracers.

\begin{figure}
\centering
\includegraphics[width=1.0\linewidth]{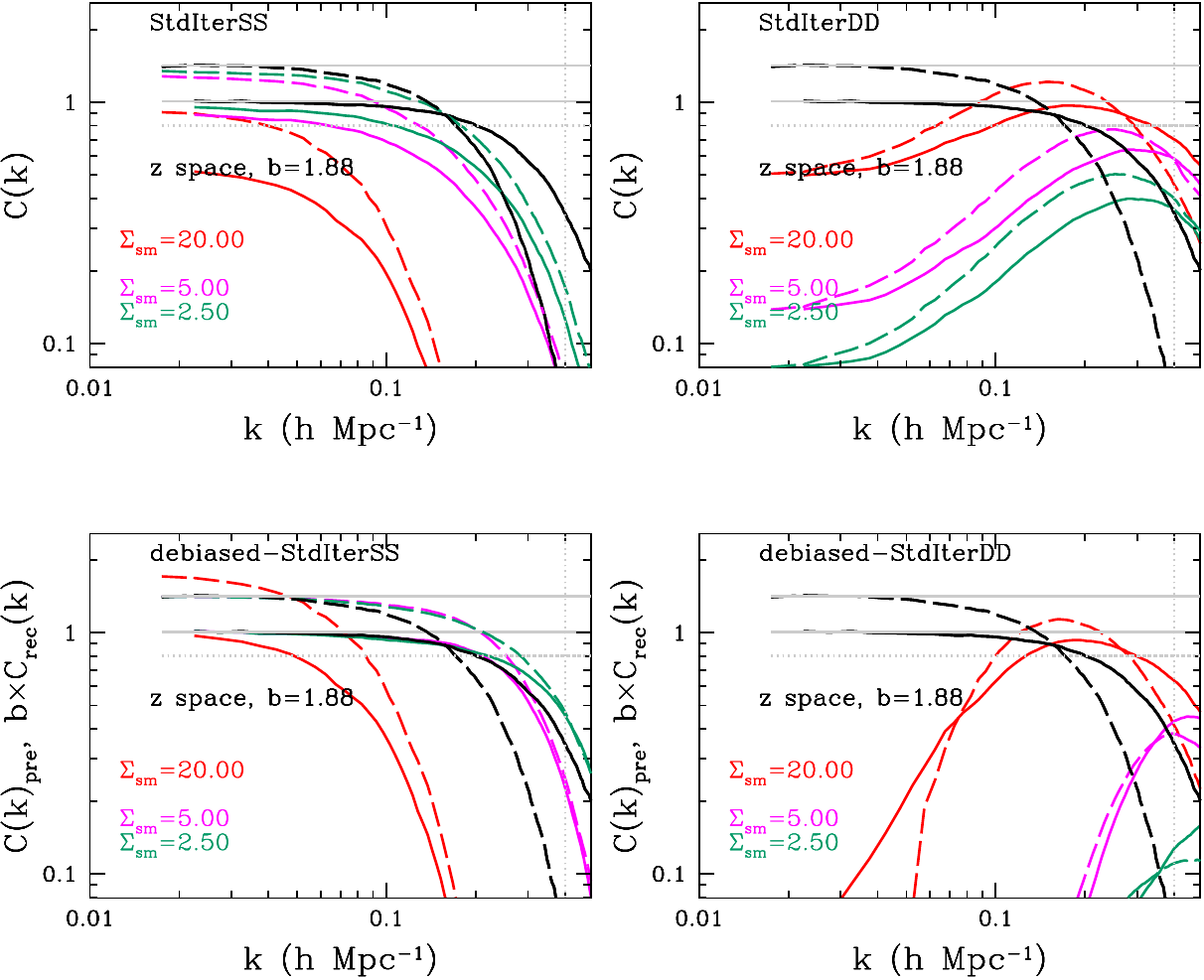}
\caption{The SS and DD component of \StdIter\ (top panels) and debiased-\StdIter\ (bottom) in Figure~\ref{fig:biased} to compare the progression of the components without and with debiasing.   Iterative reconstruction on the biased sample at $z=1$ with $b=1.88$ at various iterative steps. The quantities in the bottom panel is rescaled with $b$ to give an easy comparison to the top panels. One can see that, by debiasing, we are removing the large scale contribution from the DD component and also more effectively transferring the small scale information from DD to SS. }\label{fig:biasedssdd}
\end{figure}

The top panel of Figure \ref{fig:biased} shows $C(k)$, $P(k)$, and $R(k)$ of \StdIter\ that again started with $\Sigsm=20\hMpc$. The top panel of Figure \ref{fig:biasedssdd} shows the SS and DD components of this case.
Note that the DD component converges to $b-1$ at low $k$ after the first  reconstruction (red lines), which will incorrectly introduce a large-scale displacement in the second reconstruction, while the particles are already near their original Lagrangian space. We follow Eq.~\ref{eq:stditerbias} to reduce the large-scale displacement field at higher iterations.
Figure \ref{fig:biased} shows that, despite this inconsistency, the combination of the SS and the DD field, i.e., \StdIter\ improves the cross-correlation with iterative steps at least until the third iteration. However, this inconsistency will affect \DisIter, as this method traces the displacement itself.

We now apply {\it debiasing} described in \S~\ref{subsec:methodbias} to make sure that the standard reconstruction and \StdIter\ work with the debiasing. The middle panels of Figure~\ref{fig:biased} shows the debiased \StdIter.  As the debiasing removes the constant bias factor from the power spectrum, we rescaled $C(k)$ and $P(k)$ with $b$ and $b^2$, respectively, for these plots so that they can be directly compared to the pre-reconstruction quantities.\footnote{The nominal shot noise is subtracted from $P(k)$ after scaling with $b^2$.}
The resulting performance with and without debiasing appears consistent for \StdIter\ in terms of $R(k)$ and $C(k)$. In $P(k)$, debiasing seems to introduce a suppression in power on small scales after shot noise subtraction (magenta line), but this suppression below unity corresponds to an almost constant offset in power, as if we misestimated the shot noise contribution.
The bottom panel of Figure ~\ref{fig:biasedssdd} shows that the debiasing indeed removes the large-scale contribution of the DD component ($\delta_d$) as we intended. Here we include the reconstruction step with $\Sigsm=2.5\hMpc$ to show the behavior after the optimal step of \StdIter\ with $\Sigsm=5-10\hMpc$. Even after debiasing, probably due to the higher level of noise in the observed density field, we find that, along the transverse direction, the propagator of the SS component alone hardly is better than the nonlinear field (black) and most of the gain is the contribution from the DD component. At $\Sigsm=2.5\hMpc$, the SS component reaches its maximum, but the DD component has diminished too quickly such that the net propagator performs worse than $\Sigsm=5\hMpc$.

We note that this debiasing could be particularly useful for finding the  displacement field of the combined tracers with different halo/galaxy bias; we can estimate the underlying matter density field based on the prior knowledge of bias of the combined tracers in each spacial location and assign the debiasing weight to the reference particle. 

Next we apply \DisIter\ on the debiased field: the bottom panels of Figure~\ref{fig:biased} show the performance of \DisIter. Compared to the other two cases (top and middle panels), this \DisIter\ method seems slightly less efficient for the transverse direction, based on $R(k)$. Due to the missing small-scale power after \DisIter, the performance in $C(k)$ along the transverse direction again appears worse than the other methods. We find that if we choose a different initial smoothing scale, we can improve the performance of \DisIter. For simplicity of comparison, however, we keep the initial smoothing scale of $\Sigsm=20\hMpc$ for all biased cases. 

\begin{figure}
\centering
\includegraphics[width=1\linewidth]{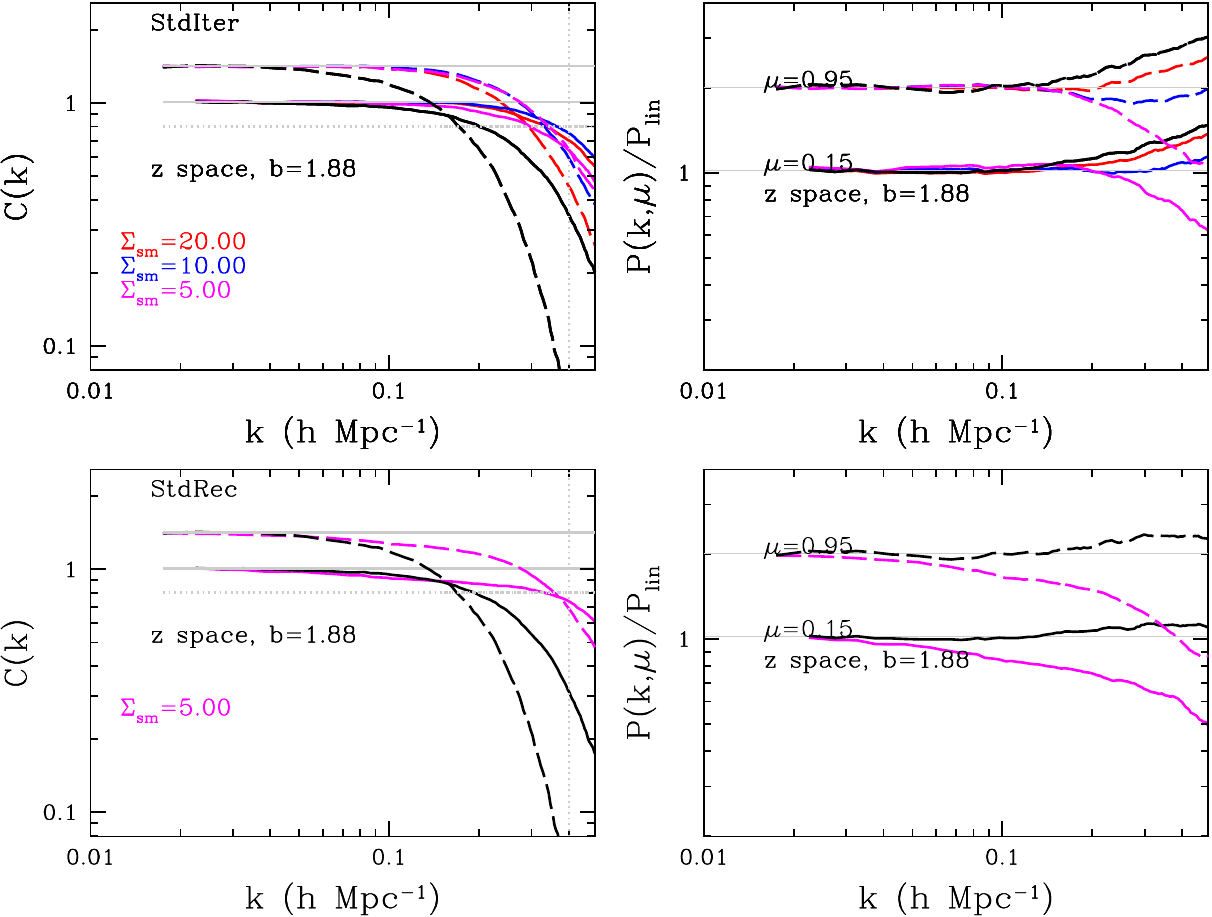}
\caption{Comparison between the iterative reconstruction \StdIter\ (top) and an example of an aggressive standard reconstruction with $\Sigsm=5\hMpc$ (bottom). In terms of the propagator at $k=0.4\ihMpc$, the standard reconstruction does not appear to fair worse than the iterative method, however, a degradation at $k\sim 0.1\ihMpc$ in terms of $P(k)$ and $C(k)$ is observed. }\label{fig:biasstdrec}
\end{figure}

Figure~\ref{fig:biasstdrec} shows the iterative reconstruction in comparison to an aggressively conducted single-step standard reconstruction for the biased cases using $b=1.88$. This again shows a result consistent with the matter case. I.e., we can find an optimal single-step smoothing scale that returns comparable $C(k)$ as the optimal iterative reconstruction, while such a single-step reconstruction with a small smoothing scale tends to show more deviation from the linear power spectrum and from the PT-based propagator model on large scales, e.g, at $k\sim 0.1-0.2\ihMpc$.

\begin{figure*}
\centering
\includegraphics[width=1\linewidth]{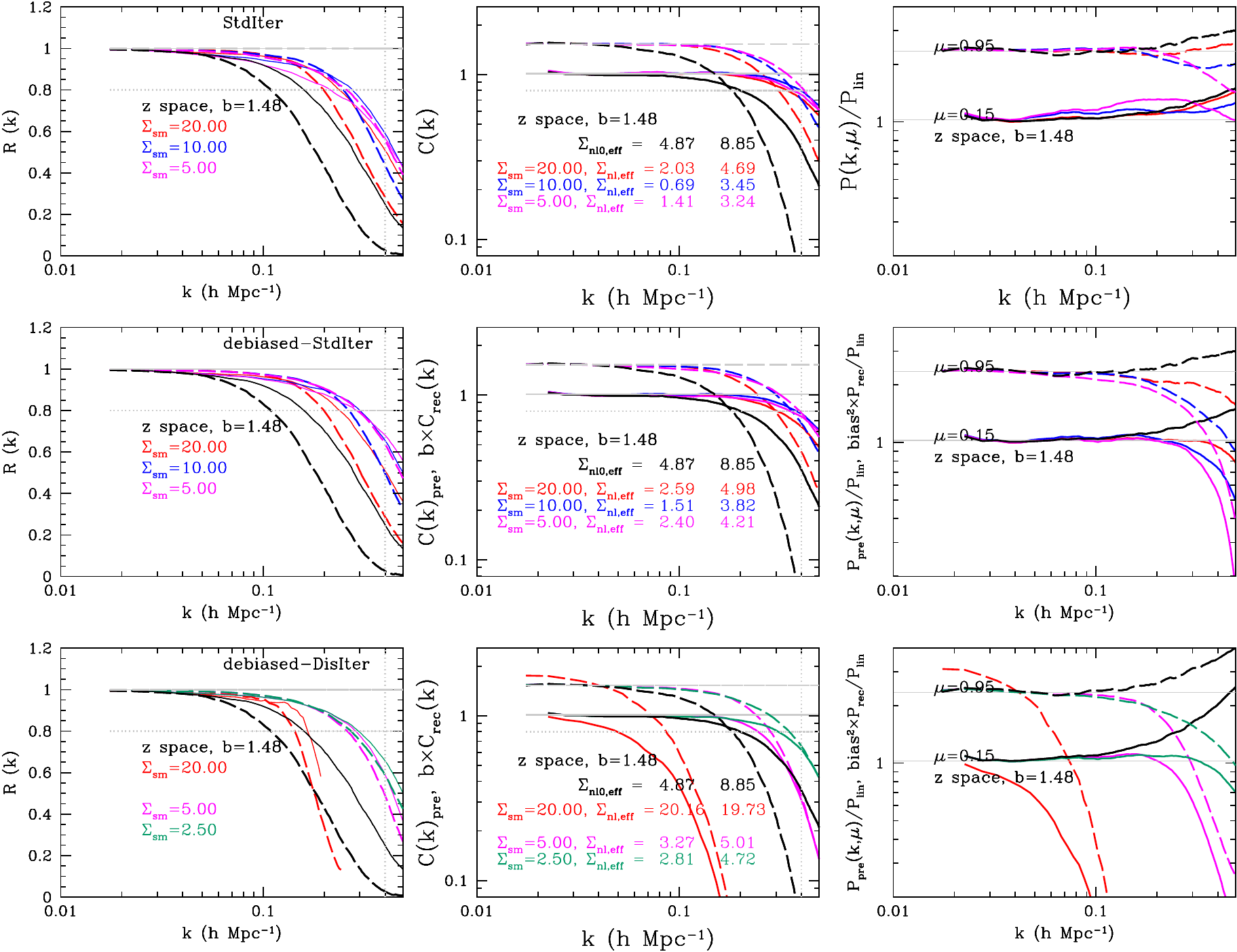}
\caption{Cross-correlation coefficient, propagator, and power spectrum of the iterative reconstruction on the biased sample at $z=1$ with $b=1.48$. Top panel: \StdIter. Middle: \StdIter\ after debiasing. Bottom: \DisIter\ after debiasing. . }\label{fig:biasedm35}
\end{figure*}

Figure~\ref{fig:biasedm35} shows the same for a lower bias and a smaller shot noise sample ($b=1.48$ and $n=0.0038\itrihMpc$). Overall, the efficiency of the reconstruction appears to improve, particularly for the \DisIter\ case with this sample, while the qualitative trend we observed from Figure~\ref{fig:biasedm35} still holds.

In summary, \StdIter\ returns a good performance even for the biased tracers despite the increased shot noise. On the other hand, we find \DisIter\ more challenging due to missing tracers. To mitigate the difficulty, we introduced debiasing. With debiasing, \DisIter\ is applicable to the biased field with realistic shot noise. In our results, the \DisIter\ option for the biased tracers appears less effective in terms of propagator along the transverse direction  when compared to \StdIter\, but this is partly because we did not try to optimize \DisIter\ for the biased field and partly because the suppression of power on small scales after \DisIter. In terms of $R(k)$, where the suppression is canceled, we see that \DisIter\ is more comparable to \StdIter. 

\section{Conclusion}\label{sec:conclusion}
The density field reconstruction technique has been widely used for recovering the BAO feature in galaxy surveys from various nonlinearities. Recently, a variety of extensions to this technique have been suggested aimed at improving the BAO information and beyond, and one main direction is to adopt the iterative steps in reconstruction, called `iterative reconstruction'. In this paper, we investigated the performance of iterative reconstruction in terms of the BAO as well as the broadband shape, focusing on the implementation based on \citet{Schmittfull:2017}.  We summarized the key results of this paper below.

\begin{itemize}
   \item We extended the methods in \citet{Schmittfull:2017} to the redshift space and tracers with halo bias and shot noise and  inspected the components of the reconstructed field in Fourier space and in configuration space. In the process of extension, we invented surrogate methods to  \citet{Schmittfull:2017} that can be applied to the galaxy field with high sparsity. 
   \item All the iterative methods we are testing yield comparable results at the low shot noise limit when redshift-space distortions and galaxy/halo bias are not included. Among these options, the displacement reconstruction method, \DisIter\, shows the best performance in terms of  the convergence in $R(k)$, $C(k)$, $P(k)$ to the linear density field.  The single-step standard reconstruction performs much worse than the final step of the iterative reconstruction, confirming previous results reported in the literature. This is expected as the effective smoothing scale is very different between the two cases. 
    \item We find that we can decrease the smoothing scale of the standard reconstruction such that its propagator (i.e.. the strength of BAO) becomes comparable to the optimal case of the iterative reconstruction. However, this can be achieved potentially at the cost of deviation from the empirical PT-based fitting model of the  BAO~\citep[e.g.,][]{Seo:2016brs} as well as from the PT-based broadband models~\citep{Hikage:2017real,Chen:2019rec}. We expect that iterative reconstruction on the other hand would allow us to use a small smoothing scale `stably', i.e., without causing a substantial deviation from the linear power spectrum and from the PT-based BAO damping model on large scales. For the dark matter example we studied, we show the deviation from the linear power spectrum at $k\sim 0.1\ihMpc$ is reduced from 13\% to 3-4\% with iterative steps. Although we have not tested explicitly in this paper, we expect that the iterative reconstruction will therefore provide a better goodness of the fit in the post-reconstruction BAO analysis for a small smoothing scale, compared to the standard reconstruction. As a caveat, our result assumed a correct linear $b$ and $f$ in the process for the biased case, at the first step reconstruction). But one can imagine that in the process of iterative operation, an fiducial, inaccurate assumption can be updated based on the reconstructed field at each step. Also, the iterative reconstruction is performed toward an almost uniform displaced galaxy density field, and this process can potentially allow a self or an internal calibration of relevant cosmological parameters (a similar point was made in ~\citet{Wang:2020}  regarding RSD parameters). We plan to investigate such aspect in a future paper. On the other hand, iterative reconstruction will require more computational time as well as a complexity in constructing a corresponding PT model for the broadband power~\citep{Ota:2021}.
   \item When redshift-space distortions are included, all iterative reconstruction methods perform worse compared to their real-space results, particularly along the line of sight. We expect that a more dedicated treatment of redshift-space distortions such as the iterative RSD correction \citep{Wang:2020} can potentially further improve the line of sight information. 
  \item  We find that the displacement-field-based reconstruction \DisIter\ becomes quickly inefficient with increasing sparsity, as a sparse field lacks the tracers of the displacement field in the majority of the pixels. We alleviate the missing tracer problem by making the reference particles to trace the reconstructed displacement (\DisIterss) and/or by debiasing. \StdIter, which is density-based, does not directly suffer the sparsity problem. 
  \item We note that the iterative reconstruction transfers the information from the galaxy field gradually to the reference fields during the iteration, returning an almost uniform galaxy field on large scales. On small scales, we observe a small, but nonzero power of the displaced galaxy density field at the last step of iteration, implying that we are not perfectly recovering the uniform Lagrangian distribution on small scales even with 9 reconstruction steps.
\end{itemize}

There are several aspects that we can improve upon the implementations made in this paper. First, we did not include a process that can enable the iterative operations to naturally converge to their optimal performance. Instead, we manually inspected and selected the optimal steps. We expect that introducing a minimum smoothing scale \citep[e.g.,][]{Schmittfull:2017}, based on the shot noise level of the raw observed field could help here.
There are also ways to improve the displacement field estimation accounting for the environment~\citep{Achitouv:2015}, weighting the halos of different mass and therefore reducing the effective shot noise \citep{Liu:2021}, and/or by accounting for the nonlinear bias \citep{Birkin:2019} which we will leave for future work. 

This paper mainly makes qualitative statements on the performance of the iterative reconstruction regarding signal-to-noise and large-scale clustering using various clustering estimators. We found the pros and cons of implementing the iterative steps and tracing the density field versus the displacement field in the presence of redshift-space distortions, halo bias, and shot noise. A more complete and quantitative comparison can be derived by investigating the properties of the covariance matrix after iterative reconstruction and an explicit BAO and broadband parameter fitting after constructing proper fitting models.  Again, we leave such a rigorous and quantitative study for future work.

\section*{Acknowledgement}
We are very grateful for the helpful comments from  Lado Samushia, Chris Blake, Martin White, and the FastPM halo catalog and comments from Zhejie Ding. H.-J.S. and A.O. are supported by the U.S.~Department of Energy, Office of Science, Office of High Energy Physics under DE-SC0019091. 
SS was supported in part by World Premier International Research Center Initiative (WPI Initiative), MEXT, Japan.

\section*{Data Availability}
The data underlying this article will be shared following any reasonable request to the corresponding author.
\bibliographystyle{mnras}
\bibliography{reference}
\end{document}